\documentclass[fleqn,10pt]{article}
\usepackage{amssymb}
\usepackage{amsmath}
\usepackage{amsfonts}
\usepackage{graphicx}

\newtheorem{theorem}{Theorem}[section]
\newtheorem{lemma}{Lemma}[section]
\newtheorem{proposition}{Proposition}[section]

\renewcommand{\theequation}{\arabic{section}.\arabic{equation}}
\newcommand{\R}{\,{\rm{Re}}\,}

\begin{document}
\title{\textbf{The Solutions of the NLS Equations with Self-Consistent Sources}}

\author{ {\bf  Yijun Shao
    \hspace{1cm} Yunbo Zeng\dag } \\
    {\small {\it
    Department of Mathematical Sciences, Tsinghua University,
    Beijing 100084, China}} \\
    {\small {\it \dag
     Email: yzeng@math.tsinghua.edu.cn}}}

\date{}
\maketitle
\renewcommand{\theequation}{\arabic{section}.\arabic{equation}}

\begin{abstract}
We construct the generalized Darboux transformation with arbitrary
functions in time $t$ for the AKNS equation with self-consistent
sources (AKNSESCS) which, in contrast with the Darboux
transformation for the AKNS equation, provides a
non-auto-B\"{a}cklund transformation between two AKNSESCSs with
different degrees of sources. The formula for N-times repeated
generalized Darboux transformation is proposed. By reduction the
generalized Darboux transformation with arbitrary functions in
time $t$ for the Nonlinear Schr\"{o}dinger equation with
self-consistent sources (NLSESCS) is obtained  and enables us to
find the dark soliton, bright soliton and positon solutions for
NLS$^{+}$ESCS and NLS$^{-}$ESCS. The properties of these solution
are analyzed.

\end{abstract}

\hskip\parindent


\section{Introduction}
\setcounter{equation}{0} \hskip\parindent

The nonlinear Schr\"{o}dinger equation with self-consistent
sources (NLSESCS) describes the Soliton propagation in a medium
with both resonant and nonresonant nonlinearities [1-4], and it
also describes the nonlinear interaction of high-frequency
electrostatic wave with ion acoustic waves in plasma
\cite{Clau91}. Some soliton solution for the NLSESCS was obtained
by inverse scattering transformation in \cite{mel92}. Since the
explicit time part of the Lax representation of the NLSESCS was
not found, the method to solve the NLSESCS by inverse scattering
transformation in \cite{mel92} was quite complicated.

Due to the important role played by the soliton
 equations with self-consistent sources (SESCSs) in many fields of
 physics, such as hydrodynamics,
solid state physics, plasma physics,  SESCSs have attracted some
attention [6-16]. In recent years we have presented method to find
the explicit time part of the  Lax representation for SESCSs and
to construct generalized binary Darboux transformations with
arbitrary functions in time $t$ for SESCSs which, in contrast with
the Darboux transformation for soliton equations \cite{Matveev91,
Manas96}, offer a non-auto-B\"{a}cklund transformation between two
SESCSs with different degrees of sources and can be used to obtain
N-soliton, positon and negaton solutions [19-21].

The positon solution for many soliton equations and their physical
application have been widely studied, for example, the positon
solutions for KdV and mKdV equations were investigated in
\cite{Matveev02, Rasi}, for the  nonlinear Schr\"{o}dinger
equation in \cite{Barran99}, for the sine-Gordon equation in
\cite{Beutler}. However positon solutions for SESCSs except for
KdV equation with self-consistent sources in [19,20] have not been
studied.

In this paper, we develop the method presented in [19,20] to study
the NLSESCS. First we construct the generalized Darboux
transformation with arbitrary functions in time $t$ for the AKNS
equation with self-consistent sources (AKNSESCS) which offers a
non-auto-B\"{a}cklund transformation between two AKNSESCSs with
different degrees of sources. Then by reduction we obtained the
generalized Darboux transformation with arbitrary functions in
time $t$ for the NLSESCS which also provides a
non-auto-B\"{a}cklund transformation between two NLSESCSs with
different degrees of sources. Some interesting solutions of
NLSESCS such as dark soliton, bright soliton and positon solutions
for NLS$^{+}$ESCS and NLS$^{-}$ESCS are found. The properties of
these solution are analyzed.

\section{Binary Darboux transformations for the AKNS equation with self-consistent sources}
\setcounter{equation}{0} \hskip\parindent The AKNSESCS is defined
as [15,16]
\begin{subequations}
\label{a1}
\begin{equation}
\label{a1a}
%\begin{array}{l}
q_{t}=-i(q_{xx}-2q^2r)+\sum_{j=1}^{n}(\varphi_j^{(1)})^2,\quad
r_{t}=i(r_{xx}-2qr^2)+\sum_{j=1}^{n}(\varphi_j^{(2)})^2,
\end{equation}
\begin{equation}
\label{a1b}
\varphi_{j,x}
=\left(\begin{array}{cc}
-\lambda_j&q\\
r&\lambda_j
\end{array}\right)
\varphi_j,\quad j=1,\cdots,n,
%\end{array}
\end{equation}
\end{subequations}
where $\lambda_j$'s are $n$ distinct complex constants,
$\varphi_j=(\varphi_j^{(1)},\varphi_j^{(2)})^T$ (hereafter, we use superscripts
$(1)$ and $(2)$ to denote the first and second element of
a two dimensional vector respectively).

The Lax pair for Eqs. (\ref{a1}) is given by [15,16]
\begin{subequations}
\label{a2}
\begin{equation}
\label{a2a}
\psi_x=U\psi,\quad U:=U(\lambda,q,r)=
\left(\begin{array}{cc}
-\lambda&q\\
r&\lambda
\end{array}\right),
\end{equation}
\begin{equation}
\label{a2b}
\psi_{t_s}=R^{(n)}\psi,\quad R^{(n)}:=V+\sum_{j=1}^n\frac{H(\varphi_j)}{\lambda-\lambda_j},
\end{equation}
\end{subequations}
where
\begin{equation*}
V:=V(\lambda,q,r)=i\left(\begin{array}{cc}
-2\lambda^2+qr&2\lambda q-q_x\\
2\lambda r+r_x&2\lambda^2-qr
\end{array}\right),
\quad
H(\varphi_j)=\frac{1}{2}
\left(\begin{array}{cc}
-\varphi^{(1)}_j\varphi^{(2)}_j&(\varphi^{(1)}_j)^2\\
-(\varphi^{(2)}_j)^2&\varphi^{(1)}_j\varphi^{(2)}_j
\end{array}\right)
\end{equation*}

\subsection{Binary Darboux transformation with an arbitrary constant}
\hskip\parindent It is known [16] that based on the Darboux
transformation for the AKNS equation [22], the AKNSESCS admits two
elementary Darboux transformations $\mathcal{T}_{1,2}:$
$(q,r,\varphi_1,\cdots,\varphi_n)$
$\mapsto$$(\widetilde{q},\widetilde{r},\widetilde{\varphi}_1,\cdots,\widetilde{\varphi}_n)$.
Given two arbitrary complex numbers $\mu$ and $\nu$, $\mu\neq\nu$,
let $f=f(\mu)$ and $g=g(\nu)$ be two solutions of (\ref{a2}) with
$\lambda=\mu$ and $\lambda=\nu$ respectively, and define\\
$\mathcal{T}_1[f]$:
\begin{equation*}
\widetilde{\psi}=T_1\psi,\quad T_1=T_1(\lambda,f)=
\left(\begin{array}{cc}
\lambda-\mu+qf^{(2)}/(2f^{(1)})&-q/2\\
-f^{(2)}/f^{(1)}&1
\end{array}\right),
\end{equation*}
\begin{equation*}
\widetilde{q}=-q_x/2-\mu q+q^2f^{(2)}/(2f^{(1)}),\quad \widetilde{r}=2f^{(2)}/f^{(1)},\quad
\widetilde{\varphi}_j=\frac{T_1(\lambda_j,f)\varphi_j}{\sqrt{\lambda_j-\mu}},
\quad j=1,\cdots,n;
\end{equation*}
$\mathcal{T}_2[g]$:
\begin{equation*}
\widetilde{\psi}=T_2\psi,\quad T_2=T_2(\lambda,g)=
\left(\begin{array}{cc}
1&-g^{(1)}/g^{(2)}\\
r/2&\lambda-\nu-rg^{(1)}/(2g^{(2)})
\end{array}\right),
\end{equation*}
\begin{equation*}
\widetilde{q}=-2g^{(1)}/g^{(2)},\quad \widetilde{r}=r_x/2-\nu r-r^2g^{(1)}/(2g^{(2)}),\quad
\widetilde{\varphi}_j=\frac{T_2(\lambda_j,g)\varphi_j}{\sqrt{\lambda_j-\nu}},
\quad j=1,\cdots,n
\end{equation*}
\begin{theorem}
The linear system (\ref{a2}) is covariant with respect to (w.r.t) the two
Darboux transformations $\mathcal{T}_1$, $\mathcal{T}_2$, i.e.,
the new variables $\widetilde{\psi}$, $\widetilde{q}$, $\widetilde{r}$ and $\widetilde{\varphi}_j$
satisfy
\begin{subequations}
\label{a3}
\begin{equation}
\label{a3a}
\widetilde{\psi}_x=\widetilde{U}\widetilde{\psi},\quad
\widetilde{U}=U(\lambda,\widetilde{q},\widetilde{r}),
\end{equation}
\begin{equation}
\label{a3b}
\widetilde{\psi}_{t}=\widetilde{R}^{(n)}\widetilde{\psi}:=
\left[V^{(s)}(\lambda,\widetilde{q},\widetilde{r})
+\sum_{j=1}^n\frac{H(\widetilde{\varphi}_j)}{\lambda-\lambda_j}\right]
\widetilde{\psi}.
\end{equation}
\end{subequations}
\end{theorem}

We now construct a new Darboux transformation based on $\mathcal{T}_1$ and $\mathcal{T}_2$.
Our method is similar to that for the KdV equation with self-consistent sources \cite{Zeng03}.
Define
\begin{equation*}
\sigma(f,g):=-\frac{W(f,g)}{2(\mu-\nu)},\quad
\sigma(f,f):=\lim_{\lambda\rightarrow\mu}\frac{-W(f(\lambda),f(\mu))}{2(\lambda-\mu)}
=\frac{1}{2}W(f,\partial_\mu f).
\end{equation*}
where $W(f,g)$ is the Wronskian $W(f,g):=f^{(1)}g^{(2)}-f^{(2)}g^{(1)}$.
We assume that we have obtained
$(\widetilde{\psi},\widetilde{q},\widetilde{r},\widetilde{\varphi}_1,\cdots,\widetilde{\varphi}_n)$
satisfying (\ref{a3}) by applying $\mathcal{T}_1[f]$ to $(\psi,q,r,\varphi_1,\cdots,\varphi_n)$.
Then we derive two linearly independent solutions of (\ref{a3}) with
$\lambda=\mu$ and in terms of $f$ only. \\
\textbf{First solution.} Let $f_1=f_1(\mu)$ ba a solution of (\ref{a2}) with
$\lambda=\mu$, and $W(f,f_1)\neq0$ (i.e., $f$ and $f_1$ are linearly independent). Then applying
$\mathcal{T}_1[f]$ to $f_1$ gives a solution of (\ref{a3}) with $\lambda=\mu$:
\begin{equation*}
\widetilde{f}_1:=T_1(\mu,f)f_1=\frac{W(f,f_1)}{2f^{(1)}}
\left(\begin{array}{c}
-q\\2\end{array}\right).
\end{equation*}
Since $W(f,f_1)$ is independent of both $x$ and $t$,
we assume $W(f,f_1)\equiv1$. Thus, we obtain the first solution of
(\ref{a3}):
\begin{equation*}
\widetilde{f}_1=\frac{1}{2f^{(1)}}
\left(\begin{array}{c}
-q\\2\end{array}\right).
\end{equation*}
\textbf{Second solution.} Note that $\psi_1(\lambda):=f(\lambda)/(\lambda-\mu)$ is a solution of (\ref{a2}).
Applying $\mathcal{T}_1[f]$ to $\psi_1$ gives a solution of (\ref{a3}):
\begin{equation*}
\widetilde{\psi}_1(\lambda)=T_1(\lambda, f)\psi_1=
\left(\begin{array}{c}
f^{(1)}(\lambda)\\0\end{array}\right)
+\frac{W(f(\mu),f(\lambda))}{2f^{(1)}(\mu)(\lambda-\mu)}
\left(\begin{array}{c}
-q\\2\end{array}\right).
\end{equation*}
Taking the limit, we find a second solution of (\ref{a3}) with
$\lambda=\mu$:
\begin{equation*}
\widetilde{f}:=\lim_{\lambda\rightarrow\mu}\widetilde{\psi}_1(\lambda)=
\left(\begin{array}{c}
f^{(1)}\\0\end{array}\right)
+\frac{\sigma(f,f)}{f^{(1)}}
\left(\begin{array}{c}
-q\\2\end{array}\right),
\end{equation*}

Let $C$ be an arbitrary constant, then the linear combination of the above solutions
\begin{equation*}
\widetilde{h}:=\widetilde{f}+2C\widetilde{f}_1=
\left(\begin{array}{c}
f^{(1)}\\0\end{array}\right)
+\frac{C+\sigma(f,f)}{f^{(1)}}
\left(\begin{array}{c}
-q\\2\end{array}\right)
\end{equation*}
is also a solution of (\ref{a3}) with $\lambda=\mu$. Apply
$\mathcal{T}_2[\widetilde{h}]$ to $(\widetilde{\psi}/(\lambda-\mu),$ $\widetilde{q},$
$\widetilde{r},$ $\widetilde{\varphi}_1,$ $\cdots,$ $\widetilde{\varphi}_n)$, i.e., define
\begin{subequations}
\label{a4}
\begin{equation}
\label{a4a}
\widehat{\psi}=T_2(\lambda,\widetilde{h})\frac{\widetilde{\psi}}{\lambda-\mu}=
\psi-\frac{f}{C+\sigma(f,f)}\sigma(f,\psi),
\end{equation}
\begin{equation}
\label{a4b}
\widehat{q}=-\frac{\widetilde{h}_1}{\widetilde{h}_2}=
q-\frac{2(f^{(1)})^2}{C+\sigma(f,f)},\quad
\widehat{r}=\frac{\widetilde{r}_x}{2}-\mu\widetilde{r}-\frac{r^2\widetilde{h}_1}{2\widetilde{h}_2}
=r-\frac{(f^{(2)})^2}{C+\sigma(f,f)}.
\end{equation}
\begin{equation}
\label{a4c}
\widehat{\varphi}_j=\frac{T_2(\lambda_j,\widetilde{h})\widetilde{\varphi}_j}{\sqrt{\lambda_j-\mu}}
=\varphi_j-\frac{f}{C+\sigma(f,f)}\sigma(f,\varphi_j),
\end{equation}
\end{subequations}
then the new variables $\widehat{\psi}$, $\widehat{q}$, $\widehat{r}$, $\widehat{\varphi}_j$
satisfy
\begin{subequations}
\label{a5}
\begin{equation}
\label{a5a}
\widehat{\psi}_x=\widehat{U}\widehat{\psi},
\end{equation}
\begin{equation}
\label{a5b}
\widehat{\psi}_{t}=\widehat{R}^{(n)}\widehat{\psi}.
\end{equation}
\end{subequations}
where $\widehat{U}=U(\lambda,\widehat{q},\widehat{r})$ and
$\widehat{R}^{(n)}=V(\lambda,\widehat{q},\widehat{r})
+\sum_{j=1}^nH(\widehat{\varphi}_j)/(\lambda-\lambda_j)$.
\begin{proposition}
Let $f$ be a solution of (\ref{a2}) with $\lambda=\mu$, and $C$ be an arbitrary
constant, then $\widehat{\psi}$, $\widehat{q}$, $\widehat{r}$ and $\widehat{\varphi_j}$
given by (\ref{a4}) present a binary Darboux transformation with an arbitrary constant
for (\ref{a2}),
%\begin{subequations}
%\label{a4}
%\begin{equation}
%\label{a4a}
%\widehat{\psi}=
%\psi-\frac{f}{C+\sigma(f,f)}\sigma(f,\psi),
%\end{equation}
%\begin{equation}
%\label{a4b}
%\widehat{q}=q-\frac{(f^{(1)})^2}{C+\sigma(f,f)},\quad
%\widehat{r}=r-\frac{(f^{(2)})^2}{C+\sigma(f,f)},
%\end{equation}
%\begin{equation}
%\label{a4c}
%\widehat{\varphi}_j=\varphi_j-\frac{f}{C+\sigma(f,f)}\sigma(f,\varphi_j),
%\quad j=1,\cdots,n,
%\end{equation}
%\end{subequations}
%The system (\ref{a2}) is covariant w.r.t. the Darboux transformation with an arbitrary
%constant (\ref{a4}),
%\begin{equation*}
%(\psi,q,r,\varphi_1,\cdots,\varphi_n)
%\mapsto(\widehat{\psi},\widehat{q},\widehat{r},\widehat{\varphi}_1,\cdots,\widehat{\varphi}_n),
%\end{equation*}
and $(\widehat{q},\widehat{r},\widehat{\varphi}_1,\cdots,\widehat{\varphi}_n)$
is a new solution of (\ref{a1}). Moreover, we have
\begin{equation*}
\widehat{q}\widehat{r}=qr-\partial_x^2\log[C+\sigma(f,f)].
\end{equation*}
\end{proposition}

\subsection{Binary Darboux transformation with an arbitrary function of $t$}
\hskip\parindent
Substituting (\ref{a4a}) into the left side of Eq. (\ref{a5b}), we have
a polynomial of $[C+\sigma(f,f)]^{-1}$:
\begin{equation*}
\widehat{\psi}_{t}=\frac{\partial}{\partial t}\left[\psi-\frac{f}{C+\sigma(f,f)}
\sigma(f,\psi)\right]=\psi_{t}-\frac{f_{t}}{C+\sigma(f,f)}\sigma(f,\psi)
-\frac{f[W(f_{t},\psi)+W(f,\psi_{t})]}{2(\mu-\lambda)[C+\sigma(f,f)]}
\end{equation*}
\begin{equation*}
+\frac{f\sigma(f,\psi)[W(f_{t},f_\mu)+W(f,f_{t,\mu})]}{2[C+\sigma(f,f)]^2}
=:\sum_{j=0}^2L_j[C+\sigma(f,f)]^{-j},
\end{equation*}
where $L_j$'s are two-dimensional vector functions defined by the last equality.
We can expect that substituting (\ref{a4}) into the right side of (\ref{a5b}) will
also give a polynomial of $[C+\sigma(f,f)]^{-1}$, but it will be more complicated.
So we just write it as
\begin{equation*}
\widehat{R}^{(n)}\widehat{\psi}=\sum_{j=0}^{3}R_j[C+\sigma(f,f)]^{-j},
\end{equation*}
where $R_j$'s are also two-dimensional vector functions dependent on $\psi$, $q$, $r$,
$\varphi_j$ and $f$ and their derivatives w.r.t. $x$. Since (\ref{a5b}) holds for
any constant $C$, we have the following lemma.
\begin{lemma}
Assume that $\psi$, $q$, $r$ and $\varphi_j$ satisfy (\ref{a2}), and let $f$ be a solution
of (\ref{a2}) with $\lambda=\mu$, then we have
\begin{equation*}
L_j=R_j,\quad j=0,1,2,\quad R_3=0,
\end{equation*}
for all $x$ and $t$.
\end{lemma}

We now replace the constant $C$ with an arbitrary function of $t$, say $c(t)$.
Since there is no derivatives w.r.t. $t$ in the expression of
$\widehat{R}^{(n)}$, if we replace $C$ with $c(t)$ in the definition of (\ref{a4}),
we will have
\begin{equation*}
\widehat{R}^{(n)}\widehat{\psi}=\sum_{j=0}^3R_j[c(t)+\sigma(f,f)]^{-j}.
\end{equation*}
But we will not have $\widehat{\psi}_{t}=\sum_{j=0}^3L_j[c(t)+\sigma(f,f)]^{-j}$
under this replacement. However, this replacement will lead to a non-auto-B\"{a}cklund
transformation.
\begin{proposition}
\label{pr1}
Let $f$ be a solution of (\ref{a2}) with $\lambda=\lambda_{n+1}$, and
$c(t)$ be an arbitrary function of $t$. If we define
\begin{subequations}
\label{b1}
\begin{equation}
\label{b1a}
\bar{\psi}=
\psi-\frac{f}{c(t)+\sigma(f,f)}\sigma(f,\psi),
\end{equation}
\begin{equation}
\label{b1b}
\bar{q}=q-\frac{(f^{(1)})^2}{c(t)+\sigma(f,f)},\quad
\bar{r}=r-\frac{(f^{(2)})^2}{c(t)+\sigma(f,f)},
\end{equation}
\begin{equation}
\label{b1c}
\bar{\varphi}_j=\varphi_j-\frac{f}{c(t)+\sigma(f,f)}\sigma(f,\varphi_j),
\quad j=1,\cdots,n,
\end{equation}
and
\begin{equation}
\label{b1d}
\bar{\varphi}_{n+1}=\frac{\sqrt{\dot{c}(t)}f}{c(t)+\sigma(f,f)}\sigma(f,\varphi_j),
\end{equation}
\end{subequations}
then the new variables
$\bar{\psi},\bar{q},\bar{r},\bar{\varphi}_1,\cdots,\bar{\varphi}_{n+1}$
satisfy a new system
\begin{subequations}
\label{b2}
\begin{equation}
\label{b2a}
\bar\psi_x=\bar U\bar\psi,\quad \bar U=U(\lambda,\bar q,\bar r),
\end{equation}
\begin{equation}
\label{b2b}
\bar\psi_{t}=\bar R^{(n+1)}\bar\psi,\quad \bar R^{(n+1)}=V(\lambda,\bar{q},\bar{r})
+\sum_{j=1}^{n+1}\frac{H(\bar{\varphi}_j)}{\lambda-\lambda_j},
\end{equation}
\end{subequations}
and
$(\bar{q},\bar{r},\bar{\varphi}_1,\cdots,\bar{\varphi}_{n+1})$ is a solution of
(\ref{a1}) with $n$ replaced by $n+1$. %degrees of freedom
%\begin{subequations}
%\label{b3}
%\begin{equation}
%\label{b3a}
%%\begin{array}{l}
%\bar q_{t}=-i(\bar q_{xx}-2\bar q^2\bar r)+\sum_{j=1}^{n+1}(\bar\varphi_j^{(1)})^2,\quad
%\bar r_{t}=i(\bar r_{xx}-2\bar q\bar r^2)+\sum_{j=1}^{n+1}(\bar\varphi_j^{(2)})^2,
%\end{equation}
%\begin{equation}
%\label{b3b}
%\bar\varphi_{j,x}=U(\lambda_j,\bar q,\bar r)\bar\varphi_j,\quad j=1,\cdots,n+1.
%%\end{array}
%\end{equation}
%\end{subequations}
Moreover, we have
\begin{equation*}
\bar q\bar r=qr-\partial_x^2\log[c(t)+\sigma(f,f)].
\end{equation*}
\end{proposition}
\textbf{Proof.}
Since no derivatives w.r.t. $t$ appear in Eq. (\ref{b2a}), it is covariant w.r.t.
the transformation defined by (\ref{b1}). Substitution of (\ref{b1a}) into the
left side of (\ref{b2b}) gives
\begin{equation*}
\bar{\psi}_{t}=\frac{\partial}{\partial t}\left[\psi-\frac{f}{c(t)+\sigma(f,f)}
\sigma(f,\psi)\right]=\psi_{t}-\frac{f_{t}}{c(t)+\sigma(f,f)}\sigma(f,\psi)
\end{equation*}
\begin{equation*}
-\frac{f[W(f_{t},\psi)+W(f,\psi_{t})]}{2(\mu-\lambda)[c(t)+\sigma(f,f)]}
+\frac{f\sigma(f,\psi)[2\dot{c}(t)+W(f_{t},f_\mu)+W(f,f_{t,\mu})]}{2[c(t)+\sigma(f,f)]^2}
\end{equation*}
\begin{equation*}
=\sum_{j=0}^2L_j[c(t)+\sigma(f,f)]^{-j}
+\frac{\dot{c}(t)f\sigma(f,\psi)}{[c(t)+\sigma(f,f)]^2}
\end{equation*}
\begin{equation*}
=\sum_{j=0}^2R_j[c(t)+\sigma(f,f)]^{-j}
+\frac{\sqrt{\dot{c}(t)}\sigma(f,\psi)}{c(t)+\sigma(f,f)}\bar\varphi^{n+1}
\end{equation*}
\begin{equation*}
=\bar R^{(n)}\bar\psi+\frac{H(\bar{\varphi}_{n+1})}{2(\lambda-\lambda_{n+1})}\bar\psi
=\bar R^{(n+1)}\bar\psi
\end{equation*}
This completes the proof.

\textbf{Example of solution.} We start from the Eqs. (\ref{a1}) with $n=0$, and the
initial solution $q=r=0$. Choose a solution of (\ref{a2}) with $n=0$, $q=r=0$ as
$f=(e^{-\lambda_1x-2i\lambda_1^2t},e^{\lambda_1x+2i\lambda_1^2t})^T$, then by Proposition
\ref{pr1}, we obtain a solution of (\ref{a1}) with $n=1$:
\begin{equation*}
q=-\frac{e^{-2\lambda_1x-4i\lambda_1^2t}}{x+4i\lambda_1t+c(t)},\quad
r=-\frac{e^{2\lambda_1x+4i\lambda_1^2t}}{x+4i\lambda_1t+c(t)},\quad
\varphi_1=\frac{\sqrt{\dot{c}(t)}}{x+4i\lambda_1t+c(t)}\left(\begin{array}{c}
e^{-\lambda_1x-2i\lambda_1^2t}\\ e^{\lambda_1x+2i\lambda_1^2t}
\end{array}\right),
\end{equation*}
where $c(t)$ is an arbitrary function.

\textbf{Remark.} The binary Darboux transformation (\ref{b1}), in fact, provides a
non-auto-B\"{a}cklund transformation between the AKNS equation with sources of
different degrees of freedom.
Since a function $c(t)$ is involved, we call it a binary Darboux transformation
with an arbitrary function of $t$. This transformation is dependent on two elements,
$c(t)$ and $f$, so we just write them together as a pair $\{c,f\}$.
\subsection{Multi-times repeated binary Darboux transformation with arbitrary functions}
\hskip\parindent It is evident that the binary Darboux
transformation with an arbitrary function can be applied $N$
times, and we will obtain the $N$-times repeated binary Darboux
transformtion with $N$ arbitrary functions.  Let $f_1,f_2,\cdots$,
be a series of solutions of (\ref{a2}) with
$\lambda=\lambda_1,\lambda_2,\cdots$, and let $c_1,c_2,\cdots$, be
a series of arbitrary functions of $t$. Let $\psi[N]$, $q[N]$,
$r[N]$, $\varphi_j[N]$ and $f_j[N]$ denote the $N$-times Darboux
transformed variables.

We define some symmetric forms. Let $c_j$ and $g_j$,
$j=1,2,\cdots$ be a series of scaler and two-dimensional vectors,
$u$ be a scaler, $h$ be a two-dimensional vector, and
$\sigma(g_i,g_j)$ and $\sigma(g_i,h)$ are defined. For
$N=1,2,\cdots$, we define five forms $W_0$, $W_1^{(i)}$ and
$W_2^{(i)}$, $i=1,2$, which are symmetric for the $N$ pairs
$\{c_j,g_j\}$, as follows:
\begin{equation*}
W_0(\{c_1,g_1\},\cdots,\{c_N,g_N\})=\det A,
\end{equation*}
\begin{equation*}
W_1^{(i)}(\{c_1,g_1\},\cdots,\{c_N,g_N\};h)=\det
\left(\begin{array}{cc} A&b\\\alpha^{(i)}&h^{(i)}, ]quad i=1,2,
\end{array}\right)
\end{equation*}
\begin{equation*}
W_2^{(i)}(\{c_1,g_1\},\cdots,\{c_N,g_N\};u)=\det
\left(\begin{array}{cc} A&(\alpha^{(i)})^T\\\alpha^{(i)}&u, \quad
i=1,2,
\end{array}\right)
\end{equation*}
where
\begin{equation*}
A=(\delta_{ij}c_i+\sigma(g_i,g_j))_{N\times N},\quad
b=(\sigma(g_1,h),\ldots,\sigma(g_N,h))^T,\quad
\alpha^{(i)}=(g_1^{(i)},\ldots,g_N^{(i)}).
\end{equation*}
For convenience, we define
\begin{equation*}
W_1(\{c_1,g_1\},\cdots,\{c_N,g_N\};h)=
\left(\begin{array}{c}
W_1^{(1)}(\{c_1,g_1\},\cdots,\{c_N,g_N\};h)\\
W_1^{(2)}(\{c_1,g_1\},\cdots,\{c_N,g_N\};h)
\end{array}\right)
\end{equation*}
\begin{lemma}
\label{le.c1}
Let $F_i[j]=\{c_i,f_i[j]\}$, $i,j=1,2,\ldots$, then for $l,k=1,2,\ldots$, we have
\begin{subequations}
\label{c1}
\begin{equation}
\label{c1a}
W_0(F_{l+1}[l],\ldots,F_{l+k}[l])
=\frac{W_0(F_l[l-1],\ldots,F_{l+k}[l-1])}{W_0(F_l[l-1])}
\end{equation}
\begin{equation}
\label{c1b}
W_1(F_{l+1}[l],\ldots,F_{l+k}[l];\psi[l])
=\frac{W_1(F_l[l-1],\ldots,F_{l+k}[l-1];\psi[l-1])}{W_0(F_l[l-1])},
\end{equation}
\begin{equation}
\label{c1c}
W_2^{(1)}(F_{l+1}[l],\ldots,F_{l+k}[l];q[l])
=\frac{W_2^{(1)}(F_l[l-1],\ldots,F_{l+k}[l-1];q[l-1])}{W_0(F_l[l-1])},
\end{equation}
\begin{equation}
\label{c1d}
W_2^{(2)}(F_{l+1}[l],\ldots,F_{l+k}[l];r[l])
=\frac{W_2^{(2)}(F_l[l-1],\ldots,F_{l+k}[l-1];r[l-1])}{W_0(F_l[l-1])},
\end{equation}
\end{subequations}
\end{lemma}
\textbf{Proof.} Let $a_{ij}=\delta_{ij}c_{l+i}+\sigma(f_{l+i}[l-1],f_{l+j}[l-1])$,
$i,j=1,2,\ldots$. Direct calculation yields
\begin{equation*}
\delta_{ij}c_{l+i}+\sigma(f_{l+i}[l],f_{l+j}[l])=a_{ij}-a_{i0}a_{00}^{-1}a_{0j}\equiv\bar a_{ij},\quad
i,j=1,2,\ldots.
\end{equation*}
Note that
\begin{equation*}
\left(\begin{array}{cccc} a_{00}&a_{01}&\cdots&a_{0k}\\
a_{10}&a_{11}&\cdots&a_{1k}\\
\vdots&\vdots&\ddots&\vdots\\
a_{k0}&a_{k1}&\cdots&a_{kk}
\end{array}\right)\left(\begin{array}{cccc}
1&-a_{00}^{-1}a_{01}&\cdots&-a_{00}^{-1}a_{0k}\\
0&1&\cdots&0\\
\vdots&\vdots&\ddots&\vdots\\
0&0&\cdots&1\end{array}\right)
=\left(\begin{array}{cccc} a_{00}&0&\cdots&0\\
a_{10}&\bar a_{11}&\cdots&\bar a_{1k}\\
\vdots&\vdots&\ddots&\vdots\\
a_{k0}&\bar a_{k1}&\cdots&\bar a_{kk}
\end{array}\right).
\end{equation*}
Taking determinant for both sides, we have
\begin{equation*}
W_0(F_l[l-1],\ldots,F_{l+k}[l-1]))=W_0(F_l[l-1])W_0(F_{l+1}[l],\ldots,F_{l+k}[l]),
\end{equation*}
which is just the eq. (\ref{c1a}). Similarly, we can prove
(\ref{c1b}), (\ref{c1c}) and (\ref{c1d}).
\begin{proposition}
For $N=1,2,3,\ldots,$ we have
\begin{subequations}
\label{c4}
\begin{equation}
\label{c4a}
\psi[N]=\frac{1}{\Delta}W_1(\{c_1,f_1\},\ldots,\{c_N,f_N\};\psi),
\end{equation}
\begin{equation}
\label{c4b}
q[N]=\frac{1}{\Delta}W_2^{(1)}(\{c_1,f_1\},\ldots,\{c_N,f_N\};q),
\end{equation}
\begin{equation}
\label{c4c}
r[N]=\frac{1}{\Delta}W_2^{(2)}(\{c_1,f_1\},\ldots,\{c_N,f_N\};r),
\end{equation}
\begin{equation}
\label{c4d}
\varphi_j[N]=\frac{1}{\Delta}W_1(\{c_1,f_1\},\ldots,\{c_N,f_N\};\varphi_j),\quad
j=1,\ldots,n,
\end{equation}
\begin{equation}
\label{c4e}
\varphi_{n+j}[N]=\frac{\sqrt{\dot{c}_j}\ }{c_j\Delta}W_1(\{c_1,f_1\},\ldots,\{c_N,f_N\};f_j),
\quad j=1,\ldots,N,
\end{equation}
and
\begin{equation}
\label{c4f}
q[N]r[N]=qr-\partial_x^2\log\Delta
\end{equation}
\end{subequations}
where $\Delta=W_0(\{c_1,f_1\},\ldots,\{c_N,f_N\})$.
\end{proposition}
\textbf{Proof.} By the definition of $\psi[N]$ and Lemma \ref{le.c1}, we have
\begin{equation*}
\psi[N]=\frac{W_1(\{c_N,f_N[N-1]\};\psi[N-1])}{W_0(\{c_N,f_N[N-1]\})}
=\frac{W_1(\{c_{N-1},f_{N-1}[N-2]\},\{c_N,f_N[N-2]\};\psi[N-2])}{W_0(\{c_{N-1},f_{N-1}[N-2]\})}
\end{equation*}
\begin{equation*}
\times\frac{W_0(\{c_{N-1},f_{N-1}[N-2]\})}{W_0(\{c_{N-1},f_{N-1}[N-2]\},\{c_N,f_N[N-2]\})}
=\cdots=\frac{W_1(\{c_1,f_1\},\ldots,\{c_N,f_N\};\psi)}{W_0(\{c_1,f_1\},\ldots,\{c_N,f_N\})},
\end{equation*}
which gives rise to the eq. (\ref{c4a}). Similarly, we can prove
(\ref{c4b}), (\ref{c4c}), (\ref{c4d}) and (\ref{c4e}).

%\textbf{Example of solution.} We start from the Eqs. (\ref{a1}) with $n=0$ and the initial
%solution $q=r=0$. Choose a solution

\section{Binary Darboux transformations for the NLS equations with self-consistent sources}
\setcounter{equation}{0} \hskip\parindent
It is well known that from the ordinary AKNS equation
\begin{equation}
\label{d1}
q_t=-i(q_{xx}-2q^2r),\quad r_t=i(r_{xx}-2qr^2).
\end{equation}
if we set $r=\varepsilon q^*$, $\varepsilon=\pm1$, then
Eqs. (\ref{d1}) are reduced to the ordinary NLS eqution
\begin{equation}
\label{d2}
q_t=i(2\varepsilon|q|^2q-q_{xx}).
\end{equation}
We call the equation with $\varepsilon=+1$ the NLS$^+$ equation and the equation with
$\varepsilon=-1$ the NLS$^-$ equation.

Similarly, we can reduce the AKNSESCS into
the NLS$^\pm$ equations with self-consistent sources (NLS$^\pm$ESCS), but the reductions
are more complicated since the sources need to be reduced as well.
First, we define two linear maps $S_+$ and $S_-$ by
\begin{equation}
\label{d5}
S_\pm: \left(\begin{array}{c}
z^{(1)}\\z^{(2)}\end{array}\right)\mapsto
\left(\begin{array}{c}
\pm z^{(2)*}\\z^{(1)*}\end{array}\right).
\end{equation}
For the reduced AKNS spectral problem, i.e., the NLS$^+$ spectral problem:
\begin{equation}
\label{d6}
\psi_x=U(\lambda,q,q^*)\psi
\end{equation}
and the NLS$^-$ spectral problem:
\begin{equation}
\label{d7}
\psi_x=U(\lambda,q,-q^*)\psi,
\end{equation}
we have the following lemma.
\begin{lemma}
{\rm(1)} If $f$ is a solution of (\ref{d6}) with $\lambda=\lambda_1$, then $S_+f$ is a solution
of (\ref{d6}) with $\lambda=-\lambda_1^*$; there exists a solution $f$ of (\ref{d6})
with $\lambda=\lambda_1$ satisfying $f^{(2)}=f^{(1)*}$  if and only if $\R\lambda_1=0$.
{\rm(2)} If $f$ is a solution of (\ref{d7}) with $\lambda=\lambda_1$, then $S_-f$ is a solution
of (\ref{d7}) with $\lambda=-\lambda_1^*$; there exists no solution $f$ of (\ref{d7}) satisfying
$f^{(2)}=f^{(1)*}$ if $q\neq0$.
\end{lemma}

The NLSESCS are reduced from
the AKNSESCS defined by
\begin{subequations}
\label{d8}
\begin{equation}
\label{d8a}
\varphi_{j,x}=U(\lambda_j,q,r)\varphi_j,\quad \varphi'_{j,x}=U(\lambda'_j,q,r)\varphi'_j,
\quad j=1,\ldots,m,
\end{equation}
\begin{equation}
\label{d8b}
\phi_{j,x}=U(\zeta_j,q,r)\phi_j,\quad j=1,\ldots,n,
\end{equation}
\begin{equation}
\label{d8c}
q_t=-i(q_{xx}-2q^2r)+\sum_{j=1}^{m}\left[(\varphi_j^{(1)})^2+({\varphi_j'}^{(1)})^2\right]+\sum_{j=1}^n(\phi_j^{(1)})^2,
\end{equation}
\begin{equation}
\label{d8d}
r_t=i(r_{xx}-2qr^2)+\sum_{j=1}^{m}\left[(\varphi_j^{(2)})^2+({\varphi_j'}^{(2)})^2\right]+\sum_{j=1}^n(\phi_j^{(2)})^2.
\end{equation}
\end{subequations}
where $\lambda_1,\ldots,\lambda_n,\lambda'_1,\ldots,\lambda'_n,\zeta_1,\ldots,\zeta_m$ are
$2n+m$ distinct constants.
The corresponding Lax pair is
\begin{equation}
\label{d9}
\psi_x=U(\lambda,q,r)\psi,\quad \psi_t=V(\lambda,q,r)\psi+\sum_{j=1}^m
\left[\frac{H(\varphi_j)}{\lambda-\lambda_j}+\frac{H(\varphi'_j)}{\lambda-\lambda'_j}\right]\psi
+\sum_{j=1}^n\frac{H(\phi_j)}{\lambda-\zeta_j}\psi.
\end{equation}

\noindent\\
\textbf{(1) Reductions to the NLS$^+$ESCS}

Let
\begin{subequations}
\label{d10}
\begin{equation}
\label{d10a}
r=q^*,\quad \lambda'_j=-\lambda_j^*\quad \varphi'_j=\pm S_+\varphi_j,\quad j=1,\ldots,m,
\end{equation}
\begin{equation}
\label{d10b} \ \R\zeta_j=0,\quad
{\phi_j^{(2)}}^*=\phi_j^{(1)}\equiv w_j,\quad j=1,\ldots,n,
\end{equation}
\end{subequations}
then Eqs. (\ref{d8}) are reduced to the NLS$^+$ESCS
\begin{subequations}
\label{d11}
\begin{equation}
\label{d11a}
\varphi_{j,x}=U(\lambda_j,q,q^*)\varphi_j,\quad j=1,\ldots,m,
\end{equation}
\begin{equation}
\label{d11b} w_{j,x}=\zeta_jw_j+qw^*_j,\ (\R\zeta_j=0),\quad
j=1,\ldots,n,
\end{equation}
\begin{equation}
\label{d11c}
q_t=i(2|q|^2q-q_{xx})+\sum_{j=1}^m\left[(\varphi_j^{(1)})^2+({\varphi_j^{(2)}}^*)^2\right]
+\sum_{j=1}^n w_k^2.
\end{equation}
\end{subequations}
And the system (\ref{d9}) is reduced to the Lax pair for the NLS$^+$ESCS
\begin{equation}
\label{d12} \psi_x=U(\lambda,q,q^*)\psi,\quad
\psi_t=V(\lambda,q,q^*)\psi+\sum_{j=1}^m
\left[\frac{H(\varphi_j)}{\lambda-\lambda_j}+\frac{H(S_+\varphi_j)}{\lambda+\lambda_j^*}\right]\psi
+\sum_{j=1}^n\frac{H((w_j,w_j^*)^T)}{\lambda-\zeta_j}\psi.
\end{equation}

\noindent\\
\textbf{(2) Reductions to the NLS$^-$ESCS}

Take $n=0$ in (\ref{d8}) and let
\begin{equation}
\label{d15}
r=-q^*,\quad \lambda'_j=-\lambda_j^*\quad \varphi'_j=\pm iS_-\varphi_j,\quad j=1,\ldots,m,
\end{equation}
then Eqs. (\ref{d8}) with $n=0$ are reduced to the NLS$^-$ESCS
\begin{subequations}
\label{d16}
\begin{equation}
\label{d16a}
\varphi_{j,x}=U(\lambda_j,q,-q^*)\varphi_j,\quad j=1,\ldots,m,
\end{equation}
\begin{equation}
\label{d16b}
q_t=i(-2|q|^2q-q_{xx})+\sum_{j=1}^m\left[(\varphi_j^{(1)})^2-(\varphi_j^{(2)*})^2\right].
\end{equation}
\end{subequations}
Correspondingly, the system (\ref{d9}) with $n=0$ is reduced to the Lax pair for
the NLS$^-$SCS
\begin{equation}
\label{d17}
\psi_x=U(\lambda,q,-q^*)\psi,\quad \psi_t=V(\lambda,q,-q^*)\psi+\sum_{j=1}^m
\left[\frac{H(\varphi_j)}{\lambda-\lambda_j}
-\frac{H(S_-\varphi_j)}{\lambda+\lambda_j^*}\right]\psi.
\end{equation}

We now reduce the Darboux transformations for the AKNSESCS
to the NLSESCS. It is easy to verify the following statements.
\begin{lemma}
{\rm (1)} Let $f$ and $g$ be two solutions of the NLS$^+$ spectral problem
$
 \psi_x=U(\lambda,q,q^*)\psi
$
with $\lambda=\mu,\nu$ respectively, and let $C$ be a complex constant
w.r.t. $x$, then we have
\begin{equation*}
\sigma(f,S_+g)^*=\sigma(S_+f,g),\qquad \sigma(S_+f,S_+g)^*=\sigma(f,g),
\end{equation*}
\begin{equation*}
\sigma(f,S_+f)^*=\sigma(S_+f,f),\qquad \sigma(S_+f,S_+f)^*=\sigma(f,f);
\end{equation*}
\begin{equation*}
W_0(\{C,f\},\{C^*,S_+f\})^*=W_0(\{C,f\},\{C^*,S_+f\}),
\end{equation*}
\begin{equation*}
W_1(\{C,f\},\{C^*,S_+f\};S_+g)^*=S_+W_1(\{C,f\},\{C^*,S_+f\};g),
\end{equation*}
\begin{equation*}
W_2^{(2)}(\{C,f\},\{C^*,S_+f\};0)^*=W_2^{(1)}(\{C,f\},\{C^*,S_+f\};0);
\end{equation*}
Moreover, if $g$ satisfies $g^{(2)}={g^{(1)}}^*$ $(\Rightarrow\R\nu=0)$, then
\begin{equation*}
W_1^{(2)}(\{C,f\},\{C^*,S_+f\};g)^*=W_1^{(1)}(\{C,f\},\{C^*,S_+f\};g).
\end{equation*}
{\rm (2)} Let $f$ and $g$ be two solutions of the NLS$^-$ spectral problem
$
\psi_x=U(\lambda,q,-q^*)\psi
$
with $\lambda=\mu,\nu$ respectively, and let $C$ is a complex constant
w.r.t. $x$, then we have
\begin{equation*}
\sigma(f,S_-g)^*=\sigma(S_-f,g),\qquad \sigma(S_-f,S_-g)^*=-\sigma(f,g),
\end{equation*}
\begin{equation*}
\sigma(f,S_-f)^*=\sigma(S_-f,f),\qquad \sigma(S_-f,S_-f)^*=-\sigma(f,f),
\end{equation*}
\begin{equation*}
W_0(\{C,f\},\{-C^*,S_-f\})^*=W_0(\{C,f\},\{-C^*,S_-f\}),
\end{equation*}
\begin{equation*}
W_1(\{C,f\},\{-C^*,S_+f\};S_-g)^*=S_-W_1(\{C,f\},\{-C^*,S_-f\};g),
\end{equation*}
\begin{equation*}
W_2^{(2)}(\{C,f\},\{-C^*,S_-f\};0)^*=-W_2^{(1)}(\{C,f\},\{-C^*,S_-f\};0).
\end{equation*}
\end{lemma}

Using this lemma, we can reduce binary Darboux transformations for the AKNSESCS
 to binary Darboux transformations for the NLSESCS.

\noindent\\
\textbf{(1) Darboux transformations for the NLS$^+$ESCS }

The binary Darboux transformation (\ref{b1}) for the AKNSSCS is reduced to a
binary Darboux transformation with an arbitrary function for the NLS$^+$ESCS as follows:
\begin{proposition}
\label{pr.e1} Given a solution
$(q,\varphi_1,\ldots,\varphi_m,w_1,\ldots,w_n)$ of the NLS$^+$ESCS (\ref{d11}),
let $c(t)$ be a real function satisfying $\dot{c}(t)\geq0$, and let $f$
be a solution of the linear system (\ref{d12}) with
$\lambda=\zeta_{n+1}$, $\R\zeta_{n+1}=0$ and satisfy $f^{(1)}=f^{(2)*}$. Define
\begin{subequations}
\label{e1}
\begin{equation}
\label{e1a}
\bar\psi=\psi-\frac{f}{c(t)+\sigma(f,f)}\sigma(f,\psi),\quad \bar
q=q-\frac{(f^{(1)})^2}{c(t)+\sigma(f,f)},
\end{equation}
\begin{equation}
\label{e1b}
\bar\varphi_j=\varphi_j-\frac{f}{c(t)+\sigma(f,f)}\sigma(f,\varphi_j),\quad
j=1,\ldots,m,
\end{equation}
\begin{equation}
\label{e1c} \bar w_j=w_j-\frac{f^{(1)}}{c(t)+\sigma(f,f)}\sigma(f,
(w_j,w_j^*)^T),\quad j=1,\ldots,n,
\end{equation}
\begin{equation}
\label{e1d} \bar
w_{n+1}=\frac{\sqrt{\dot{c}(t)}\,f^{(1)}}{c(t)+\sigma(f,f)},
\end{equation}
\end{subequations}
then the new variables $\bar\psi$, $\bar q$,
$\bar\varphi_1,\ldots,\bar\varphi_m$ and $\bar w_1,\ldots,\bar
w_{n+1}$ satisfy the system (\ref{d12}) with $n$ replaced by
$n+1$. Hence $(\bar q,\bar\varphi_1,\ldots,\bar\varphi_n,\bar
w_1,\ldots,\bar w_{m+1})$ is a solution of the NLS$^+$ESCS
(\ref{d11}) with $n$ replaced by $n+1$. Moreover, we have
\begin{equation}
\label{e3}
|\bar q|^2=|q|^2-\partial_x^2\log[c(t)+\sigma(f,f)].
\end{equation}
\end{proposition}

The two times repeated binary Darboux transformation for the AKNSESCS can be
reduced to a second binary Darboux transformation with an arbitrary function
for the NLS$^+$ESCS as follows:
\begin{proposition}
Given a solution $(q,\varphi_1,\ldots,\varphi_m,w_1,\ldots,w_n)$
of the NLS$^+$ESCS (\ref{d11}), let $c(t)$ be
an arbitrary complex function, and $f$ be a solution of the linear
system (\ref{d12}) with $\lambda=\lambda_{m+1}$,
$\R\lambda_{m+1}\neq0$. Let $\Delta=W_0(\{c,f\},\{c^*,S_+f\})$,
and define
\begin{subequations}
\label{e4}
\begin{equation}
\label{e4a}
\bar\psi=\Delta^{-1}W_1(\{c,f\},\{c^*,S_+f\};\psi),
\end{equation}
\begin{equation}
\label{e4b}
\bar q=q+\Delta^{-1}W_2^{(1)}(\{c,f\},\{c^*,S_+f\};0),
\end{equation}
\begin{equation}
\label{e4c}
\bar\varphi_j=\Delta^{-1}W_1(\{c,f\},\{c^*,S_+f\};\varphi_j),\quad
j=1,\ldots,m
\end{equation}
\begin{equation}
\label{e4d} \bar
w_j=\Delta^{-1}W_1^{-1}(\{c,f\},\{c^*,S_+f\};(w_j,w_j^*)^T), \quad
j=1,\ldots,n,
\end{equation}
\begin{equation}
\label{e4e}
\bar\varphi_{m+1}=\sqrt{\dot{c}}\,(c\Delta)^{-1}W_1(\{c,f\},\{c^*,S_+f\};f),
\end{equation}
\end{subequations}
then the new variables $\bar\psi$, $\bar q$,
$\bar\varphi_1,\ldots,\bar\varphi_{m+1}$ and $\bar w_1,\ldots,\bar
w_{n}$ satisfy the system (\ref{d12}) with $m$ replaced by $m+1$.
Hence $(\bar q,\bar\varphi_1,\ldots,\bar\varphi_{n+1},\bar
w_1,\ldots,\bar w_{m})$ is a solution of the NLS$^+$ESCS
(\ref{d11}) with $m$ replaced by $m+1$. Moreover, we have
\begin{equation}
\label{e6}
|\bar q|^2=|q|^2-\partial_x^2\log\Delta.
\end{equation}
\end{proposition}

If we repeat Darboux transformation (\ref{e1}) for $N$ times and Darboux transformation
(\ref{e4}) for $M$ times, then we have a general multi-times repeated Darboux transformation
with $N+M$ arbitrary functions as follows:
\begin{proposition}
\label{pr.e3} Given a solution
$(q,\varphi_1,\ldots,\varphi_m,w_1,\ldots,w_n)$ of the NLS$^+$ESCS (\ref{d11}),
let $f_j$ be a solution of the
linear system (\ref{d12}) with $\lambda=\zeta_{n+j}$,
$\R\zeta_{n+j}=0$, and satisfy $f_j^{(1)}=f_j^{(2)*}$, $j=1,\ldots,N$,
and let $g_j$ be a solution of the
linear system (\ref{d12}) with $\lambda=\lambda_{m+j}$, $\R\lambda_{m+j}\neq0$,
$j=1,\ldots,M$. Let $c_j(t)$ be an arbitrary
real function satisfying $\dot{c}_j(t)\geq0$, $j=1,\ldots,N$, and let $d_j(t)$ be an arbitrary
complex function, $j=1,\ldots,M$. Let $F_j=\{c_j,f_j\}$,
$G_j=\{d_j,g_j\}$, $G'_k=\{d_k^*,S_+g_j\}$, and
$\Delta=W_0(F_1,\ldots,F_N,G_1,G'_1,\ldots,G_M,G'_M)$, and
define
\begin{subequations}
\label{e7}
\begin{equation}
\label{e7a}
\bar\psi=\Delta^{-1}W_1(F_1,\ldots,F_N,G_1,G'_1,\ldots,G_M,G'_M;\psi),
\end{equation}
\begin{equation}
\label{e7b}
\bar q=q+\Delta^{-1}W_2^{(1)}(F_1,\ldots,F_N,G_1,G'_1,\ldots,G_M,G'_M;0),
\end{equation}
\begin{equation}
\label{e7c}
\bar\varphi_j=\Delta^{-1}W_1(F_1,\ldots,F_N,G_1,G'_1,\ldots,G_M,G'_M;\varphi_j),
\quad j=1,\ldots,m,
\end{equation}
\begin{equation}
\label{e7d}
\bar\varphi_{m+j}=\sqrt{\dot{c}_j}\,(c_j\Delta)^{-1}
W_1(F_1,\ldots,F_N,G_1,G'_1,\ldots,G_M,G'_M;g_j),
\quad j=1,\ldots,M,
\end{equation}
\begin{equation}
\label{e7e} \bar
w_j=\Delta^{-1}W_1^{(1)}(F_1,\ldots,F_N,G_1,G'_1,\ldots,G_M,G'_M;(w_j,w_j^*)^T),
\quad j=1,\ldots,n,
\end{equation}
\begin{equation}
\label{e7f} \bar w_{n+j}=\sqrt{\dot{d}_j}\,(d_j\Delta)^{-1}
W_1(F_1,\ldots,F_N,G_1,G'_1,\ldots,G_M,G'_M;f_j), \quad
j=1,\ldots,N,
\end{equation}
\end{subequations}
then the new variables $\bar\psi$, $\bar q$,
$\bar\varphi_1,\ldots,\bar\varphi_{m+M}$ and $\bar w_1,\ldots,\bar
w_{n+N}$ satisfy the system (\ref{d12}) with $m,n$ replaced by
$m+M,n+N$, respectively. Hence $(\bar
q,\bar\varphi_1,\ldots,\bar\varphi_{m+M},\bar w_1,\ldots,\bar
w_{n+N})$ is a solution of the NLS$^+$ESCS (\ref{d11}) with $m,n$
replaced by $m+M,n+N$. Moreover, we have
\begin{equation}
\label{e9}
|\bar q|^2=|q|^2-\partial_x^2\log\Delta.
\end{equation}
\end{proposition}
\noindent
\textbf{(2) Darboux transformations for the NLS$^-$ESCS}

The binary Darboux transformation for the AKNSESCS cannot be reduced to a Darboux
transformation for the NLS$^-$ESCS. But the two-times Darboux transformation for
the AKNSESCS can be reduced to a binary Darboux transformation with an arbitrary
function for the NLS$^-$ESCS.
\begin{proposition}
\label{pr.e4}
Given a solution $(q,\varphi_1,\ldots,\varphi_m)$ of the NLS$^-$ESCS (\ref{d16}),
let $f$ be a solution of the linear system
(\ref{d17}) with $\lambda=\lambda_{m+1}$, $\R\lambda_{m+1}\neq0$.
Let $c(t)$ be an arbitrary complex function, $\Delta=W_0(\{c,f\},\{-c^*,S_-f\})$, and define
\begin{subequations}
\label{e10}
\begin{equation}
\label{e10a}
\bar\psi=\Delta^{-1}W_1(\{c,f\},\{-c^*,S_-f\};\psi),
\end{equation}
\begin{equation}
\label{e10b}
\bar q=q+\Delta^{-1}W_2^{(1)}(\{c,f\},\{-c^*,S_-f\};0),
\end{equation}
\begin{equation}
\label{e10c}
\bar\varphi_j=\Delta^{-1}W_1(\{c,f\},\{-c^*,S_-f\};\varphi_j),\quad
j=1,\ldots,m
\end{equation}
\begin{equation}
\label{e10d}
\bar\varphi_{n+1}=\sqrt{\dot{c}}\,(c\Delta)^{-1}W_1(\{c,f\},\{-c^*,S_-f\};f),
\end{equation}
\end{subequations}
then the new variables $\bar\psi$, $\bar q$, $\bar\varphi_1,\ldots,\bar\varphi_{m+1}$
 satisfy the system (\ref{d17}) with $m$ replaced by $m+1$,
Moreover, we have
\begin{equation}
\label{e12}
|\bar q|^2=|q|^2+\partial_x^2\log\Delta.
\end{equation}
\end{proposition}

Repeating the above Darboux transformation for $N$ times gives rise to a general $N$-times
repeated binary Darboux transformation with $N$ arbitrary functions for the NLS$^-$ESCS.
\begin{proposition}
Given a solution $(q,\varphi_1,\ldots,\varphi_m)$ of the NLS$^-$ equations
with sources (\ref{d16}),
let $f_j$ be a solution of the linear system
(\ref{d17}) with $\lambda=\lambda_{m+j}$, $\R\lambda_{m+j}\neq0$, $j=1,\ldots,N$.
Let $c_j(t)$ be an arbitrary complex function, $F_j=\{c_j,f_j\}$, $F'_j=\{-c_j^*,S_-f_j\}$,
$j=1,\ldots,N$, $\Delta=W_0(F_1,F'_1,\ldots,F_N,F'_N)$, and define
\begin{subequations}
\label{e13}
\begin{equation}
\label{e13a}
\bar\psi=\Delta^{-1}W_1(F_1,F'_1,\ldots,F_N,F'_N;\psi),
\end{equation}
\begin{equation}
\label{e13b}
\bar q=q+\Delta^{-1}W_2^{(1)}(F_1,F'_1,\ldots,F_N,F'_N;0),
\end{equation}
\begin{equation}
\label{e13c}
\bar\varphi_j=\Delta^{-1}W_1(F_1,F'_1,\ldots,F_N,F'_N;\varphi_j),\quad
j=1,\ldots,m
\end{equation}
\begin{equation}
\label{e13d}
\bar\varphi_{m+j}=\sqrt{\dot{c}_j}\,(c_j\Delta)^{-1}W_1(F_1,F'_1,\ldots,F_N,F'_N;f_j),
\quad j=1,\ldots,N
\end{equation}
\end{subequations}
then the new variables $\bar\psi$, $\bar q$, $\bar\varphi_1,\ldots,\bar\varphi_{m+1}$
 satisfy the system (\ref{d17}) with $m$ replaced by $m+N$,
and hence $(\bar q,\bar\varphi_1,\ldots,\bar\varphi_{m+N})$ is a
solution of the NLS$^+$ESCS (\ref{d16}) with $m$ replaced by
$m+N$. Moreover, we have
\begin{equation}
\label{e15}
|\bar q|^2=|q|^2+\partial_x^2\log\Delta.
\end{equation}
\end{proposition}

\section{Solutions of the NLS equations with sources}
\setcounter{equation}{0} \hskip\parindent
This section is devoted to the obtaining some examples of the solutions of the NLSESCS
by Darboux transformations and the analysis for these solutions.
We use subscripts $z_R$ and $z_I$ to indicate the real part and the imaginary part of a
complex number $z$.
For $\forall\,z=|z|e^{i\theta}\in\mathbb{C}$ with $\theta\in(-\pi,\pi]$, we define
$\sqrt{z}=\sqrt{|z|}\,e^{i\theta/2}$.

\subsection{Solutions of the NLS$^+$ESCS}
\hskip\parindent
We only consider the NLS$^+$ESCS (\ref{d11}) with $m=0$.
We start from the NLS$^+$ESCS (i.e., $m=n=0$)
\begin{equation}
\label{f1}
q_t=i(2|q|^2q-q_{xx})
\end{equation}
and its solution
\begin{equation}
\label{f2}
q=\rho e^{2i\rho^2t},
\end{equation}
where $\rho\in\mathbb{R}_+$ is a constant. We need to solve the linear system
\begin{equation}
\label{f3}
\psi_x=U(\lambda,\rho e^{2i\rho^2t},\rho e^{-2i\rho^2t})\psi,
\quad \psi_t=V(\lambda,\rho e^{2i\rho^2t},\rho e^{-2i\rho^2t})\psi.
\end{equation}
The fundamental solution matrix for the linear system (\ref{f4}) is
\begin{equation}
\label{f4}
\Psi=\left(\begin{array}{cc}
\rho e^{\kappa(x+2i\lambda t)+i\rho^2t}&(\kappa+\lambda)e^{-\kappa(x+2i\lambda t)+i\rho^2t}\\
(\kappa+\lambda)e^{\kappa(x+2i\lambda t)-i\rho^2t}&-\rho e^{-\kappa(x+2i\lambda t)-i\rho^2t}
\end{array}\right),
\end{equation}
where $\kappa=\kappa(\lambda)$ satisfies $\kappa^2=\lambda^2+\rho^2$.

\subsubsection{Solutions of the NLS$^+$ESCS with $m=0$ and $n=1$.}
\hskip\parindent
The NLS$^+$ESCS with $m=0$ and $n=1$ reads
\begin{subequations}
\label{f5}
\begin{equation}
\label{f5a}
w_{1,x}=i\ell w_1+qw_1^{*},
\end{equation}
\begin{equation}
\label{f5b}
q_t=i(2|q|^2q-q_{xx})+w_1^2.
\end{equation}
\end{subequations}
where $\ell\neq0$ is a real constant. Let $f$ be a solution of the system (\ref{f3})
with $\lambda=i\ell$ and satisfy $f^{(1)}=f^{(2)*}$, and let $c(t)$ be an arbitrary real
function with $\dot{c}(t)\geq0$, then by Proposition \ref{pr.e3}, a solution to the equation is given by
\begin{equation}
\label{f6}
q=\rho e^{2i\rho^2t}-\frac{(f_1^{(1)})^2}{c(t)+\sigma(f,f)},
\quad
w_1=\frac{\sqrt{\dot{c}(t)}\,f_1^{(1)}}{c(t)+\sigma(f,f)}.
\end{equation}
Moreover, we have
\begin{equation}
\label{f7}
|q|^2=\rho^2-\partial_x^2\log[c(t)+\sigma(f,f)].
\end{equation}
For the two cases: $\rho>|\ell|$ and $\rho<|\ell|$, formulas (\ref{f6}) will give two different classes
of solutions respectively: dark one-soliton solution and one-positon solution.

\noindent\\
\textbf{(1) Dark one-soliton solution and scattering property.}

We take $\rho>|\ell|$ and let
$\kappa_1=\kappa(i\ell)$.
We choose $\kappa=\sqrt{\lambda^2+\rho^2}$, then $\kappa$ and $\sqrt{\kappa\pm\lambda}$ are
analytic at $\lambda=i\ell$, and $\kappa_1=\sqrt{\rho^2-\ell^2}>0$.
Taking into account that the equality $\rho=\sqrt{\kappa-\lambda}\sqrt{\kappa+\lambda}$
holds near $\lambda=i\ell$,
we choose $f$ as
\begin{equation*}
f=\left[\Psi\left(\begin{array}{c}
\sqrt{\kappa-\lambda}\,/\rho\\0
\end{array}\right)\right]_{\lambda=i\ell}
=\left.\left(\begin{array}{c}
\sqrt{\kappa-\lambda}\,e^{\kappa(x+2i\lambda t)+i\rho^2t}\\
\sqrt{\kappa+\lambda}\,e^{\kappa(x+2i\lambda t)-i\rho^2t}
\end{array}\right)\right|_{\lambda=i\ell}
=\left(\begin{array}{c}
\sqrt{\kappa_1-i\ell}\,e^{\kappa_1(x-2\ell t)+i\rho^2t}\\
\sqrt{\kappa_1+i\ell}\,e^{\kappa_1(x-2\ell t)-i\rho^2t}
\end{array}\right).
\end{equation*}
Then one finds that $f^{(2)}=f^{(1)*}$. Calculation yields
\begin{equation*}
\sigma(f,f)=\frac{1}{2}\left|\begin{array}{cc}
f^{(1)}&\partial_{(i\ell)}f^{(1)}\\
f^{(2)}&\partial_{(i\ell)}f^{(2)}
\end{array}\right|
%=\left|\begin{array}{cc}
%\sqrt{\kappa_1-\zeta_1}&\sqrt{\kappa_1-\zeta_1}\,(\frac{d(\kappa\xi)}{d\lambda}|_{\lambda=\zeta_1}
%-(2\kappa_1)^{-1})\\
%\sqrt{\kappa_1+\zeta_1}&\sqrt{\kappa_1+\zeta_1}\,(\frac{d(\kappa\xi)}{d\lambda}|_{\lambda=\zeta_1}
%+(2\kappa_1)^{-1})
%\end{array}\right|e^{2\kappa_1\xi_1}
=\frac{\rho}{2\kappa_1}e^{2\kappa_1(x-2\ell t)}.
\end{equation*}
Let $c(t)=(2\kappa_1)^{-1}\rho e^{2\kappa_1(at+b)}$ with $a\in\mathbb{R}_+$, $b\in\mathbb{R}$
being constants,
then formulas (\ref{f6}) give a dark one-soliton solution
\begin{subequations}
\label{f8}
\begin{equation}
\label{f8a}
q=\rho e^{2i\rho^2t}-\frac{2\kappa_1(\kappa_1-i\ell)e^{2\kappa_1(x-2\ell t)+2i\rho^2t}}
{\rho(e^{2\kappa_1(at+b)}+e^{2\kappa_1(x-2\ell t)})}=
\frac{1-e^{-4i\theta}e^{2\xi}}{1+e^{2\xi}}\rho e^{2i\rho^2t},
\end{equation}
\begin{equation}
\label{f8b}
w_1=\sqrt{\frac{a(\kappa_1-i\ell)}{\rho}}\,\frac{2\kappa_1e^{\kappa_1(x-2\ell
t)+\kappa_1(at+b)+i\rho^2t}}
{e^{2\kappa_1(at+b)}+e^{2\kappa_1(x-2\ell t)}}=
\frac{2\sqrt{a}\,\kappa_1e^{\xi-i\theta}}
{1+e^{2\xi}}e^{i\rho^2t},
\end{equation}
\end{subequations}
where
\begin{equation*}
\label{f9}
\xi=\kappa_1[x-(2\ell+a)t-b],\quad \theta=\frac{1}{2}\arcsin\frac{\ell}{\rho}.
\end{equation*}
By formula (\ref{f7}), one obtains
\begin{equation}
\label{f10}
|q|^2=\rho^2-\partial_x^2\log(1+e^{2\xi})=\rho^2-\frac{\kappa_1^2}{\cosh^2\xi},
\end{equation}
which shows that $|q|^2$ describes the propagation of a dark
soliton on the constant background $\rho$. The soliton is localized around $\xi=0$,
so the location
of the soliton is $x(t)=(2\ell+a)t+b$. and the the soliton velocity is
$2\ell+a$. If $a=0$,
then $w_1\equiv0$, and $q$ defined by (\ref{f8}) becomes a dark one-soliton
solution \cite{Faddeev87} of the NLS$^+$ equation (\ref{f1}).

We fix a solution of the system (\ref{f3}) as
\begin{equation}
\label{f11}
\psi_0(x,t;\lambda)=\left(\begin{array}{c}
\rho e^{i\rho^2t}\\(\kappa+\lambda)e^{-i\rho^2t}
\end{array}\right)e^{\kappa(x+2i\lambda t)},
\end{equation}
Then a solution of the NLS$^+$ spectral problem
\begin{equation}
\label{f12}
\psi_x=U(\lambda,q,q^*)\psi%\quad \bar\psi_t=V(\lambda,\bar q,\bar q^*)\bar\psi
%+\frac{H((\bar w_1,\bar w_1^*)^T)}{\lambda-i\ell}\bar\psi,
\end{equation}
with $q$ defined by (\ref{f8}) is given by
\begin{equation*}
\psi=\psi_0-\frac{f\sigma(f,\psi_0)}{c(t)+\sigma(f,f)}
=\left(\begin{array}{c}
\rho e^{i\rho^2t}\\(\kappa+\lambda)e^{-i\rho^2t}
\end{array}\right)e^{\kappa(x+2i\lambda t)}-
\left(\begin{array}{c}
\sqrt{\kappa_1-i\ell}\,e^{i\rho^2t}\\\sqrt{\kappa_1+i\ell}\,e^{-i\rho^2t}
\end{array}\right)\frac{\kappa_1e^{2\xi}e^{\kappa(x+2i\lambda t)}}
{\rho(\lambda-i\ell)(1+e^{2\xi})}
\times
\end{equation*}
\begin{equation}
\label{f13}
\times\left|\begin{array}{cc}
\sqrt{\kappa_1-i\ell}&\rho\\
\sqrt{\kappa_1+i\ell}&\kappa+\lambda
\end{array}\right|
=\frac{(\rho^2+i\ell\lambda-\kappa_1\kappa)e^{2\xi}}{\rho^2(\lambda-i\ell)(1+e^{2\xi})}
\left(\begin{array}{c}
\rho(\kappa_1-i\ell)e^{i\rho^2t}\\-(\kappa+\lambda)(\kappa_1+i\ell)e^{-i\rho^2t}
\end{array}\right)e^{\kappa(x+2i\lambda t)}.
\end{equation}

Based on formulas (\ref{f8}), we can analyze the asymptotic features of the dark
one-soliton solution. For fixed $t$, we have
\begin{equation}
\label{f14}
q=\left\{\begin{array}{ll}
\rho e^{2i\rho^2t}[1+o(1)],&x\rightarrow-\infty,\\
\rho e^{i(\pi-4\theta)}e^{2i\rho^2t}[1+o(1)],&x\rightarrow+\infty,
\end{array}\right.
\end{equation}
\begin{equation}
\label{f15}
w_1\rightarrow 0,\quad x\rightarrow\pm\infty.
\end{equation}
% We see that the dark one-soliton solution of the NLS$^+$
%equation with a source and that of the NLS$^+$ equation have the same
%shape and the same asymptotic property as $x\rightarrow\infty$. The only difference
%between them is the velocity of the soliton, which is confined in the interval
%$(-2\rho,2\rho)$ for the latter, while takes value in $(-2\rho,+\infty)$ for the former since
%$a$ is an arbitrary positive real number.
It is easy to see that $q$ belongs to the class of potentials satisfying the
finite density boundary condition \cite{Faddeev87}
\begin{equation}
\label{f16}
q(x,t)=\rho e^{i\alpha_\pm(t)}[1+o(1)],\quad x\rightarrow\pm\infty,
\end{equation}
where $\alpha_\pm(t)$ are real functions and $\beta\equiv\frac{1}{2}(\alpha_+(t)-\alpha_-(t))$
is a real constant independent of $t$. We now define the scattering data
for this class of potentials in a similar way in \cite{Matveev02}.

First, we define $u=q e^{-i\alpha_-(t)}$, then $u$ satisfies the standard finite density
 boundary condition
\begin{equation}
\label{f17}
u(x,t)=\left\{\begin{array}{ll}
\rho[1+o(1)],&x\rightarrow-\infty,\\
\rho e^{2i\beta}[1+o(1)],&x\rightarrow+\infty.
\end{array}\right.
\end{equation}
Next, we define transmission and reflection coefficients for the NLS$^+$ spectral system
\begin{equation}
\label{f23}
\phi_x=\left(\begin{array}{cc}
-\lambda&u\\u^*&\lambda
\end{array}\right)\phi.
\end{equation}
For $u\equiv\rho$, the system (\ref{f3}) has two linearly independent solutions
\begin{equation*}
\left(\begin{array}{c}
\frac{\rho}{\kappa+\lambda}\\1
\end{array}\right)e^{\kappa x},\quad
\left(\begin{array}{c}
-1\\\frac{\rho}{\kappa+\lambda}
\end{array}\right)e^{-\kappa x},
\end{equation*}
while for $u\equiv\rho e^{2i\beta}$, the system (\ref{f12}) has two linearly independent solutions
\begin{equation*}
Q(\beta)\left(\begin{array}{c}
\frac{\rho}{\kappa+\lambda}\\1
\end{array}\right)e^{\kappa x},\quad
Q(\beta)\left(\begin{array}{c}
-1\\\frac{\rho}{\kappa+\lambda}
\end{array}\right)e^{-\kappa x},
\end{equation*}
where $Q(\beta)={\rm diag}(e^{i\beta},e^{-i\beta})$. We fix a Jost solution $\phi$
of the system (\ref{f12}) by
imposing the asymptotic property
\begin{equation}
\label{f24}
\phi=\left(\begin{array}{c}
\frac{\rho}{\kappa+\lambda}\\1
\end{array}\right)e^{\kappa x}[1+o(1)],\quad x\rightarrow-\infty,
\end{equation}
while the transmission and reflection coefficients $a(\lambda,t)$ and $b(\lambda,t)$ are
determined by the asymptotic estimate
\begin{equation}
\label{f25}
\phi= a(\lambda,t)Q(\beta)\left(\begin{array}{c}
\frac{\rho}{\kappa+\lambda}\\1
\end{array}\right)e^{\kappa x}
+b(\lambda,t)Q(\beta)\left(\begin{array}{c}
-1\\\frac{\rho}{\kappa+\lambda}
\end{array}\right)e^{-\kappa x},\quad x\rightarrow+\infty.
\end{equation}

We can now calculate the scattering data for the dark one-soliton solution. In this case,
we have $u=qe^{-i\rho^2t}$ and $\beta=\pi/2-2\theta$. formula (\ref{f13}) implies the
the function $\psi$ has the asymptotic behaviors
\begin{equation}
\label{f26}
\psi=\left(\begin{array}{c}
\rho e^{i\rho^2t}\\(\kappa+\lambda)e^{-i\rho^2t}
\end{array}\right)e^{\kappa(x+2i\lambda t)}[1+o(1)],\quad x\rightarrow-\infty,
\end{equation}
\begin{equation}
\label{f27}
\psi=\frac{\rho^2+i\ell\lambda-\kappa_1\kappa}{\rho^2(\lambda-i\ell)}
\left(\begin{array}{c}
\rho(\kappa_1-i\ell)e^{i\rho^2t}\\-(\kappa+\lambda)(\kappa_1+i\ell)e^{-i\rho^2t}
\end{array}\right)e^{\kappa(x+2i\lambda t)}[1+o(1)],\quad x\rightarrow+\infty,
\end{equation}
We now take the Jost solution
\begin{equation}
\label{f28}
\phi=Q(-\rho^2t)(\kappa+\lambda)^{-1}e^{-2i\kappa\lambda t}\psi,
\end{equation}
then we have
\begin{equation}
\label{f29}
\phi=\frac{\rho^2+i\ell\lambda-\kappa_1\kappa}
{i\rho(\lambda-i\ell)}Q(\pi/2-2\theta)
\left(\begin{array}{c}
\frac{\rho}{\kappa+\lambda}\\1
\end{array}\right)e^{\kappa x}[1+o(1)],\quad x\rightarrow+\infty,
\end{equation}
which implies that
\begin{equation}
\label{f30}
a(\lambda,t)=\frac{\rho^2+i\ell\lambda-\kappa_1\kappa}
{i\rho(\lambda-i\ell)},\quad b(\lambda,t)=0.
\end{equation}
The dark one-soliton solution is a reflectionless potential.

\noindent\\
\textbf{(2) One-positon solution and super-reflectionless property}

We take $\rho<|\ell|$ and choose
$\kappa=({\rm sign}\,\lambda_I)\,i\sqrt{-\lambda^2-\rho^2}$, then
$\kappa$ is analytic at $\lambda=i\ell$, and $\kappa(i\ell)=ik_1$,
where $k_1=({\rm sign}\,\ell)\sqrt{\ell^2-\rho^2}$ is a real constant.
Choose a periodic solution of the system (\ref{f3}) with $\lambda=i\ell$ as
\begin{equation*}
f=\left[\Psi\left(\begin{array}{c}
1\\-1
\end{array}\right)\right]_{\lambda=i\ell}
=\left.\left(\begin{array}{c}
\rho e^{\kappa(x+2i\lambda t)+i\rho^2t}-(\kappa+\lambda)e^{-\kappa(x+2i\lambda t)+i\rho^2t}\\
(\kappa+\lambda)e^{\kappa(x+2i\lambda t)-i\rho^2t}+\rho e^{-\kappa(x+2i\lambda t)-i\rho^2t}
\end{array}\right)\right|_{\lambda=i\ell}
\end{equation*}
\begin{equation}
\label{f31}
=\left(\begin{array}{c}
\rho  e^{i(\Theta+\rho^2t)}-i(k_1+\ell)e^{-i(\Theta-\rho^2t)}\\
i(k_1+\ell)e^{i(\Theta-\rho^2t)}+\rho e^{-i(\Theta+\rho^2t)}
\end{array}\right),
\end{equation}
where $\Theta=k_1(x-2\ell t)$. One finds
$f^{(2)}=f^{(1)*}$, and
\begin{equation*}
(f^{(1)})^2=2\ell(k_1+\ell)[-k_1\ell^{-1}\cos2\Theta+i(\sin2\Theta-\rho\ell^{-1})]e^{2i\rho^2t},
\end{equation*}
\begin{equation*}
\sigma(f,f)=2\ell(k_1+\ell)[x-2(k_1^2\ell^{-1}+\ell)t+\rho(2k_1\ell)^{-1}\cos2\Theta].
\end{equation*}

Choose $c(t)=2\ell(k_1+\ell)(at+b)$ with $a\in\mathbb{R}_+$, $b\in\mathbb{R}$
being constants, which implies that $\dot{c}(t)\geq0$.
Then formulas (\ref{f6}) give a one-positon solution
\begin{subequations}
\label{f32}
\begin{equation}
q=\rho e^{2i\rho^2t}-\frac{(f^{(1)})^2}{c(t)+\sigma(f,f)}=\left[\rho+
\frac{k_1\ell^{-1}\cos2\Theta-i(\sin2\Theta-\rho\ell^{-1})}
{\gamma+\rho(2k_1\ell)^{-1}\cos2\Theta}\right]e^{2i\rho^2t},
\end{equation}
\begin{equation}
w_1=\frac{\sqrt{\dot{c}(t)}\,f^{(1)}}{c(t)+\sigma(f,f)}=\sqrt{\frac{a}{2}}\cdot
\frac{\sqrt{1-k_1\ell^{-1}}\,e^{i\Theta}-i\sqrt{1+k_1\ell^{-1}}\,e^{-i\Theta}}
{\gamma+(2k_1\ell)^{-1}\rho\cos2\Theta}\,e^{i\rho^2t},
\end{equation}
\end{subequations}
where
\begin{equation*}
\gamma=x+[a-2(\ell+k_1^2\ell^{-1})]t+b.
\end{equation*}
Formula (\ref{f7}) implies
\begin{equation}
\label{f33}
|q|^2=\rho^2-\partial_x^2\log[\gamma+(2k_1\ell)^{-1}\rho\cos2\Theta]
=\rho^2+\frac{1+\rho^2\ell^{-2}+2\rho\ell^{-1}(k_1\gamma\cos2\Theta-\sin2\Theta)}
{[\gamma+\rho(2k_1\ell)^{-1}\cos2\Theta]^2}.
\end{equation}
When $\rho=a=0$, we have $k_1=\ell$ and $w_1\equiv0$, and formulas (\ref{f32}) degenerate to
a solution of the NLS$^+$ equation (\ref{f1})
\begin{equation}
\label{f35}
q=-\frac{e^{-2i\ell(x-2\ell t)}}{x-4\ell t+b},
%\quad w_1=-\frac{i\sqrt{a}\,e^{-i\ell(x-2\ell t)}}{x+(a-4\ell)t+b}.
\end{equation}
which was given in \cite{Barran99}.

A solution of the NLS$^+$ spectral problem (\ref{f12}) with the potential $q$ defined
by (\ref{f32}) is
\begin{equation*}
\psi=\psi_0-\frac{f\sigma(f,\psi_0)}{c(t)+\sigma(f,f)}=
\left(\begin{array}{c}
\rho e^{i\rho^2t}\\(\kappa+\lambda)e^{-i\rho^2t}
\end{array}\right)e^{\kappa(x+2i\lambda t)}
-\left(\begin{array}{c}
[\rho e^{i\Theta}-i(k_1+\ell)e^{-i\Theta}]e^{i\rho^2t}\\
{[i(k_1+\ell)e^{i\Theta}+\rho e^{-i\Theta}]}e^{-i\rho^2t}
\end{array}\right)\times
\end{equation*}
\begin{equation}
\label{f34}
\times\frac{e^{\kappa(x+2i\lambda t)}}{4\ell(\lambda-i\ell)(k_1+\ell)
[\gamma+(2k_1\ell)^{-1}\rho\cos2\Theta]}
\left|\begin{array}{cc}
\rho e^{i\Theta}-i(k_1+\ell)e^{-i\Theta}&\rho\\
i(k_1+\ell)e^{i\Theta}+\rho e^{-i\Theta}&\kappa+\lambda
\end{array}\right|
\end{equation}

Based on formulas (\ref{f32}) and (\ref{f34}), we can analyze the
basic features of the one-positon solution. Formulas (\ref{f32})
imply that for fixed $t$ and $x\rightarrow\pm\infty$, we have the
asymptotic estimate
\begin{equation}
\label{f36}
qe^{-2i\rho^2t}=\rho+[k_1\ell^{-1}\cos2\Theta-i(\sin2\Theta-\rho\ell^{-1})]x^{-1}
[1+O(x^{-1})],
\end{equation}
\begin{equation}
\label{f37}
w_1e^{-i\rho^2t}=\sqrt{a/2}(\sqrt{1-k_1\ell^{-1}}\,e^{i\Theta}
-i\sqrt{1+k_1\ell^{-1}}\,e^{-i\Theta})
x^{-1}[1+O(x^{-1})]
\end{equation}
for all $\rho\in\mathbb{R}_+$. However, the asymptotic behavior of $|q|^2$
for $\rho=0$ is different with that for $\rho>0$. Actually, for $\rho=0$, we have
\begin{equation*}
|q|^2=x^{-2}[1+O(x^{-1})],
\end{equation*}
while for $\rho>0$, we have
\begin{equation}
\label{f38}
|q|^2=\rho^2+2k_1\rho\ell^{-1} x^{-1}\cos2\Theta[1+O(x^{-1})].
\end{equation}
Compared to the dark one-soliton solution, the one-positon solution converges to its
background slowly.

As a function of $x$, the potential $q$ and the source $w_1$ share the
same first-order pole $x=x_0(t)$, which is
implicitly determined by the equation
\begin{equation*}
2k_1\ell[x_0+(a-2\ell-2k_1^2\ell^{-1})t+b]=\rho\cos(2k_1x_0-4k_1\ell t).
\end{equation*}
The uniqueness of the solution $x_0$ can be easily proved.
Let $\gamma_0(t)=x_0(t)+(a-2\ell-2k_1^2\ell^{-1})t+b$, then $\gamma_0$ satisfies
\begin{equation*}
2k_1\ell\gamma_0=\rho\cos(2k_1[\gamma_0-(a-2k_1^2\ell^{-1})t-b]).
\end{equation*}
This equation implies that $\gamma_0(t)$ is a periodic function of $t$ with period
$\ell\pi/(2k_1^3)$.
We define the velocity of a positon as the velocity of its pole.
%because a positon is concentrated around its pole.
From this definition, the
velocity of the positon is
\begin{equation*}
\label{f8}
v(t)=v(t+T)=\dot{x}_0(t)=[2\ell+2k_1^2\ell^{-1}-a+\dot{\gamma}_0(t)],
\end{equation*}
where $T=\pi/(2k_1^3)$,
and the average speed of the positon is
\begin{equation*}
\label{f9}
\frac{1}{T}\int_0^T v(t)dt=(2\ell+2k_1^2\ell^{-1}-a).
\end{equation*}

In Figure 1, we plot a one-positon solution of the NLS$^+$ESCS (\ref{f5}).
\begin{figure}[htp]
\centering
\includegraphics[width=2in,angle=270]{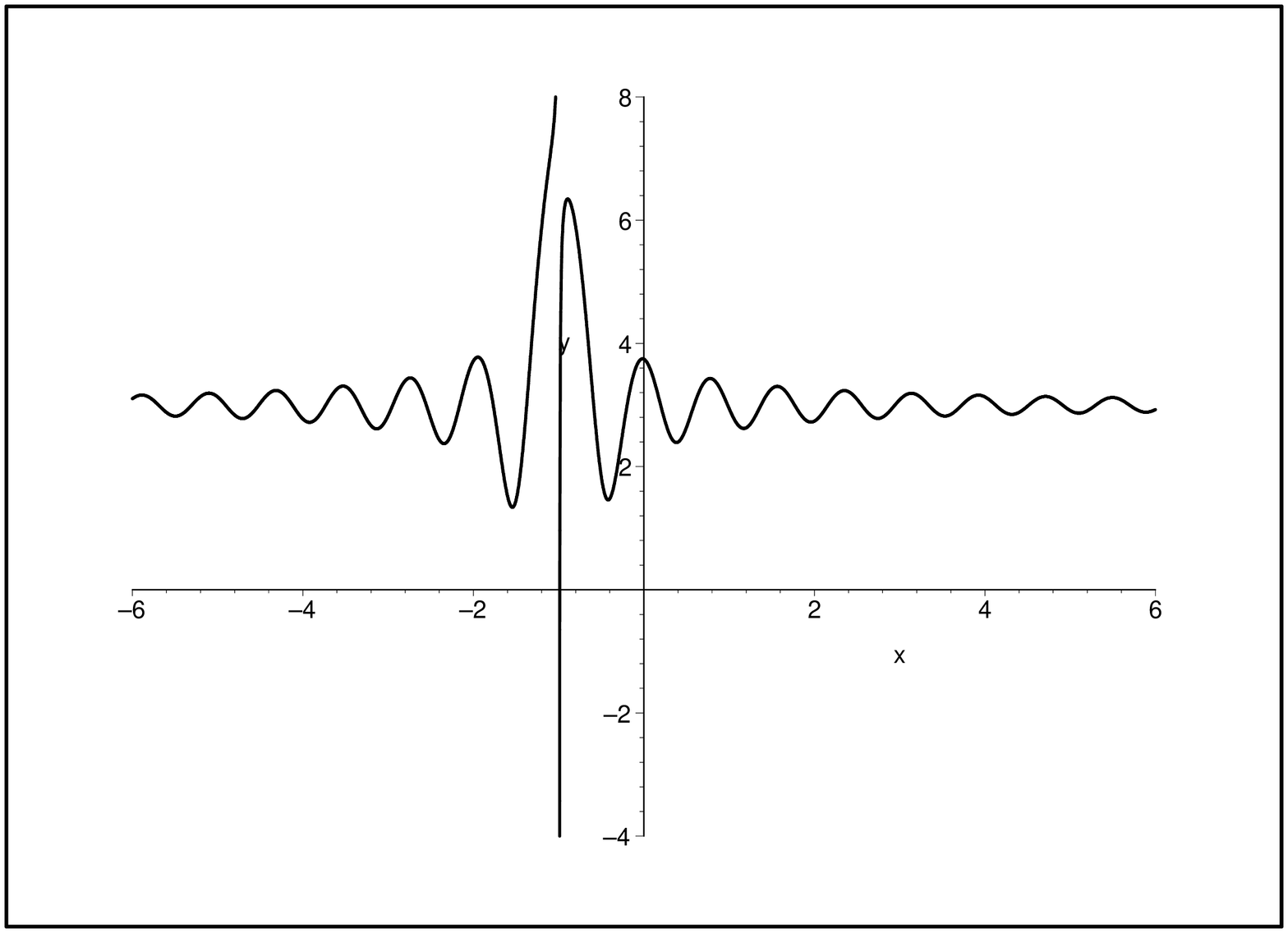}
\includegraphics[width=2in,angle=270]{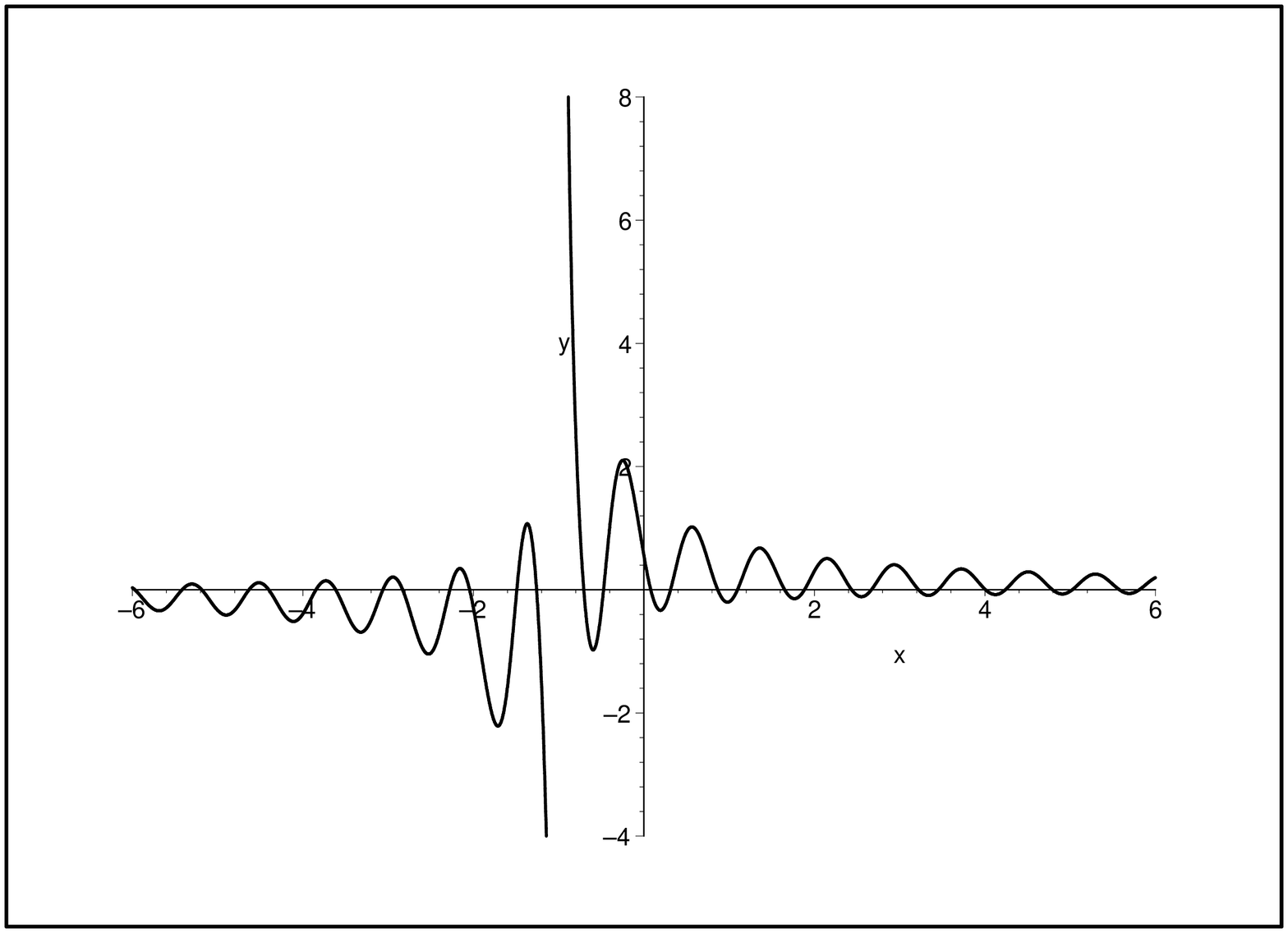}
\includegraphics[width=2in,angle=270]{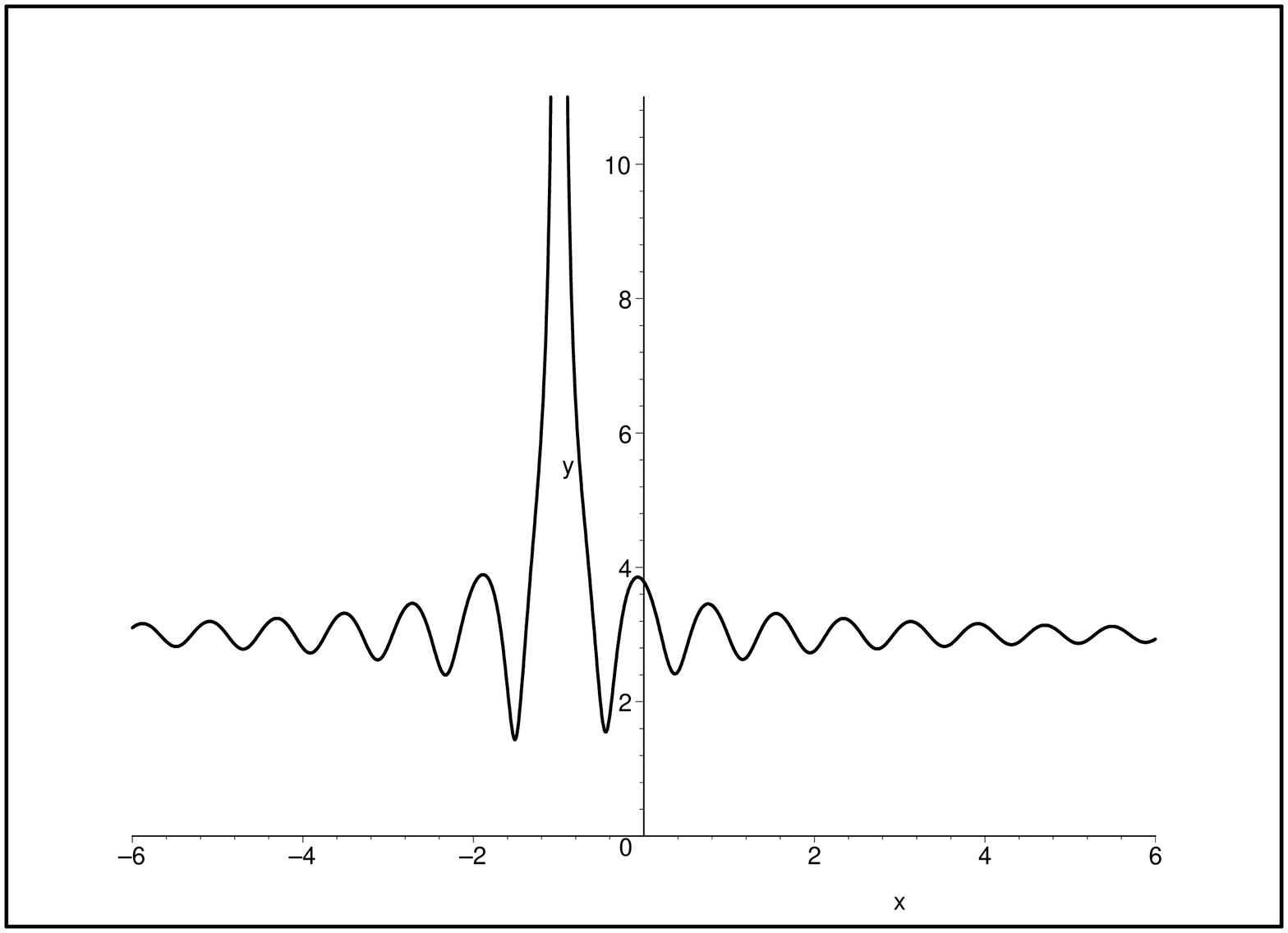}
\includegraphics[width=2in,angle=270]{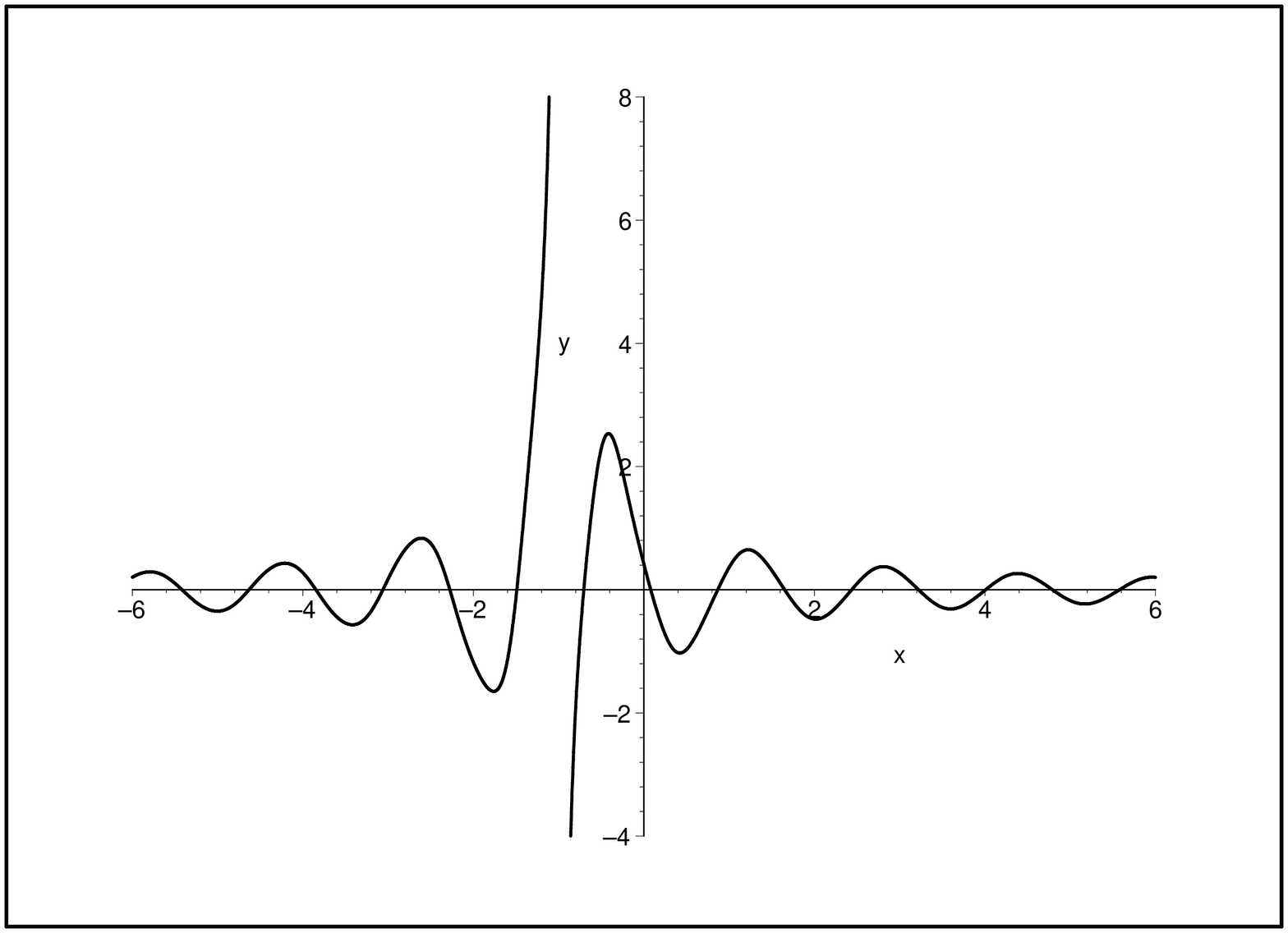}
\caption{{\footnotesize The one-positon solution of the NLS$^+$ESCS (\ref{f5}) with $\ell=5$.
The data is $\rho=3$, $a=2$ and $b=1$. The plots are taken at $t=0$. The two upper graphs
show the real and imaginary parts of $q$ respectivly while the two lower graphs
show the modulus of $q$ and the real part of $w_1$ respectivly.}}
\end{figure}

We now calculate the scattering data for the one-positon solution (\ref{f5}). In this case,
$u=qe^{-i\rho^2t}$ and $\beta=0$.
Formula (\ref{f34}) implies the asymptotic behavior of the function $\psi$
\begin{equation*}
\psi=\left(\begin{array}{c}
\rho e^{i\rho^2t}\\(\kappa+\lambda)e^{-i\rho^2t}
\end{array}\right)e^{\kappa(x+2i\lambda t)}[1+o(1)],\quad
x\rightarrow\pm\infty.
\end{equation*}
We take the Jost solution as
\begin{equation*}
\phi=Q(-\rho^2t)(\kappa+\lambda)^{-1}e^{-2i\kappa\lambda t}\psi,
\end{equation*}
then we have
\begin{equation*}
\phi\rightarrow\left(\begin{array}{c}
\frac{\rho}{\kappa+\lambda}\\1
\end{array}\right)e^{\kappa x},\quad
x\rightarrow\pm\infty,\quad
%\end{equation*}
%which implies that
%\begin{equation*}
a(\lambda,t)=1,\quad b(\lambda,t)=0.
\end{equation*}
Potentials with reflection coefficient $b=0$ and transmission
coefficient $a=1$  are called superreflectionless or
supertransparent potentials \cite{Matveev02}. By this definition,
the one-positon solution is superreflectionless.

In \cite{Matveev02}, positons are defined as long-range analogous of solitons and slowly decreasing,
oscillating solutions of nonlinear integrable equations.
If we stick to the property of slowly decreasing,
the potential $q$ defined by (\ref{f32})
should not be called a one-positon solution unless $\rho=0$. However, we see that other
properties such as long-range analogous of a soliton, super-reflectionless property
are still valid. Thus it is reasonable to extend the definition of positons as:
long-range analogous of solitons, slowly converging, oscillating solutions
of nonlinear integrable equations. According to this extended definition, the solution
(\ref{f32}) is a positon solution.

\subsubsection{Solutions of the NLS$^+$ equation with sources with $m=0$ and $n=2$.}
\hskip\parindent
The NLS$^+$ equation with sources with $m=0$ and $n=2$ reads
\begin{subequations}
\label{f39}
\begin{equation}
w_{1,x}=i\ell_1w_1+q w^*_1,\quad
w_{2,x}=i\ell_2w_2+q w^*_2,
\end{equation}
\begin{equation}
q_t=i(2|q|^2q-q_{xx})+w_1^2+w_2^2.
\end{equation}
\end{subequations}
where $\ell_1$ and $\ell_2$ are two distinct real constants. For $j=1,2$,
let $f_j$ be a solution of the system (\ref{f4}) with $\lambda=i\ell_j$ and  satisfy
$f_j^{(1)}=f_j^{(2)*}$, and let $c_j(t)$ be an arbitrary function with $\dot{c}_j(t)\geq0$.
 Then by Proposition
\ref{pr.e3}, a solution of the Eqs. (\ref{f39}) is given by
\begin{subequations}
\label{f40}
\begin{equation}
\label{f40a}
q=\rho e^{2i\rho^2t}+\frac{2\sigma(f_1,f_2)f_1^{(1)}f_2^{(1)}
-(c_1(t)+\sigma(f_2,f_2))(f_2^{(1)})^2-(c_2(t)+\sigma(f_1,f_1))(f_1^{(1)})^2}
{(c_1(t)+\sigma(f_1,f_1))(c_2(t)+\sigma(f_2,f_2))-\sigma(f_1,f_2)^2}
\end{equation}
\begin{equation}
\label{f40b}
w_1=\frac{\sqrt{\dot{c}_1(t)}\,[(c_2(t)+\sigma(f_2,f_2))f_1^{(1)}-\sigma(f_1,f_2)f_2^{(1)}]}
{(c_1(t)+\sigma(f_1,f_1))(c_2(t)+\sigma(f_2,f_2))-\sigma(f_1,f_2)^2},
\end{equation}
\begin{equation}
\label{f40c}
w_2=\frac{\sqrt{\dot{c}_2(t)}\,[(c_1(t)+\sigma(f_1,f_1))f_2^{(1)}-\sigma(f_1,f_2)f_1^{(1)}]}
{(c_1(t)+\sigma(f_1,f_1))(c_2(t)+\sigma(f_2,f_2))-\sigma(f_1,f_2)^2}.
\end{equation}
\end{subequations}
Moreover, we have
\begin{equation}
\label{f41}
|q|^2=\rho^2-\partial_x^2\log[(c_1(t)+\sigma(f_1,f_1))(c_2(t)
+\sigma(f_2,f_2))-\sigma(f_1,f_2)^2].
\end{equation}
For simplicity, we assume $|\ell_1|>|\ell_2|$. According to the three cases for $\rho$:
(i) $\rho>|\ell_j|$, $j=1,2$, (ii) $\rho<|\ell_j|$, $j=1,2$, and (iii)
$|\ell_1|>\rho>|\ell_2|$, formulas (\ref{f40}) will give three classes of solutions
respectively: dark two-soliton solution, two-positon solution and
one-soliton-one-positon solution.

\noindent\\
\textbf{(1) Dark two-soliton solution.}

For $j=1,2$, we take $\rho>|\ell_j|$, and choose
\begin{equation*}
f_j=\left[\Psi\left(\begin{array}{c}
\sqrt{\kappa-\lambda}\,/\rho\\0
\end{array}\right)\right]_{\lambda=i\ell_j}
=\left(\begin{array}{c}
\sqrt{\kappa_j-i\ell_j}\,e^{\kappa_j(x-2\ell_jt)+i\rho^2t}\\
\sqrt{\kappa_j+i\ell_j}\,e^{\kappa_j(x-2\ell_jt)-i\rho^2t}
\end{array}\right),
\end{equation*}
where $\kappa=\sqrt{\lambda^2+\rho^2}$ and $\kappa_j=\sqrt{\rho^2-\ell_j^2}$.
Let
\begin{equation*}
 c_j(t)=\frac{\rho}{2\kappa_j}e^{2\kappa_j(a_jt+b_j)}
\quad \theta_j=\frac{1}{2}\arcsin\frac{\ell_j}{\rho},\quad j=1,2,
\end{equation*}
where $a_j\in\mathbb{R}_+$, $b_j\in\mathbb{R}$ are constants,
then one finds
\begin{equation*}
\sigma(f_j,f_j)=\frac{\rho}{2\kappa_j}e^{2\kappa_j(x-2\ell_jt)},\quad j=1,2,\quad
\sigma(f_1,f_2)=\frac{\rho\sin(\theta_1-\theta_2)}
{\ell_1-\ell_2}e^{\kappa_1(x-2\ell_1t)+\kappa_2(x-2\ell_2t)},.
\end{equation*}
Formulas (\ref{f40}) yield a dark two-soliton solution
\begin{subequations}
\label{f42}
\begin{equation*}
q=\frac{1}{\Delta}\left|\begin{array}{ccc}
\frac{\rho}{2\kappa_1}(1+e^{2\xi_1})&
\frac{\rho\sin(\theta_1-\theta_2)}{\ell_1-\ell_2}e^{\xi_1+\xi_2}&
\sqrt{\rho}\,e^{\xi_1+i(\rho^2t-\theta_1)}\\
\frac{\rho\sin(\theta_1-\theta_2)}{\ell_1-\ell_2}e^{\xi_1+\xi_2}&
\frac{\rho}{2\kappa_2}(1+e^{2\xi_2})&
\sqrt{\rho}\,e^{\xi_2+i(\rho^2t-\theta_2)}\\
\sqrt{\rho}\,e^{\xi_1+i(\rho^2t-\theta_1)}&
\sqrt{\rho}\,e^{\xi_2+i(\rho^2t-\theta_2)}&\rho e^{2i\rho^2t}
\end{array}\right|
\end{equation*}
\begin{equation}
=\frac{\rho e^{2i\rho^2t}}{\Delta}\left|\begin{array}{ccc}
\frac{\rho}{2\kappa_1}(1+e^{2\xi_1})&
\frac{\rho\sin(\theta_1-\theta_2)}{\ell_1-\ell_2}e^{\xi_1+\xi_2}&
e^{\xi_1-i\theta_1}\\
\frac{\rho\sin(\theta_1-\theta_2)}{\ell_1-\ell_2}e^{\xi_1+\xi_2}&
\frac{\rho}{2\kappa_2}(1+e^{2\xi_2})&
e^{\xi_2-i\theta_2}\\
e^{\xi_1-i\theta_1}&
e^{\xi_2-i\theta_2}&1
\end{array}\right|,
\end{equation}
\begin{equation}
w_1=\frac{\sqrt{a_1}\,\rho
e^{i\rho^2t}}{\Delta}\left|\begin{array}{cc}
\frac{\rho}{2\kappa_2}(1+e^{2\xi_2})&
\frac{\rho\sin(\theta_1-\theta_2)}{\ell_1-\ell_2}e^{\xi_1+\xi_2}\\
e^{\xi_2-i\theta_2}&
e^{\xi_1-i\theta_1}
\end{array}\right|,
\end{equation}
\begin{equation}
w_2=\frac{\sqrt{a_2}\,\rho
e^{i\rho^2t}}{\Delta}\left|\begin{array}{cc}
\frac{\rho}{2\kappa_1}(1+e^{2\xi_1})&
\frac{\rho\sin(\theta_1-\theta_2)}{\ell_1-\ell_2}e^{\xi_1+\xi_2}\\
e^{\xi_1-i\theta_1}&
e^{\xi_2-i\theta_2}
\end{array}\right|,
\end{equation}
\end{subequations}
where $\xi_j=\kappa_j[x-(2\ell_j+a_j)t-b_j]$, $j=1,2$, and
\begin{equation*}
\Delta=\left|\begin{array}{cc}
\frac{\rho}{2\kappa_1}(1+e^{2\xi_1})&
\frac{\rho\sin(\theta_1-\theta_2)}{\ell_1-\ell_2}e^{\xi_1+\xi_2}\\
\frac{\rho\sin(\theta_1-\theta_2)}{\ell_1-\ell_2}e^{\xi_1+\xi_2}&
\frac{\rho}{2\kappa_2}(1+e^{2\xi_2})
\end{array}\right|.
\end{equation*}

\noindent\\
\textbf{(2) Two-positon solutions and positon-positon interaction}

For $j=1,2$, we take $\rho<|\ell_j|$, and choose
\begin{equation*}
f_j=\left[\Psi\left(\begin{array}{c}
1\\-1\\
\end{array}\right)\right]_{\lambda=i\ell_j}=
\left(\begin{array}{c}
[\rho e^{i\Theta_j}-i(k_j+\ell_j)e^{-i\Theta_j}]e^{i\rho^2t}\\
{[i(k_j+\ell_j)e^{i\Theta_j}+\rho e^{-i\Theta_j}]}e^{-i\rho^2t}
\end{array}\right),
\end{equation*}
where $\kappa=({\rm sign}\,\lambda_I)\,i\sqrt{-\lambda^2-\rho^2}$, and
$\Theta_j=k_j(x-2\ell_jt)$, $k_j=({\rm sign}\,\ell_j)\sqrt{\ell_j^2-\rho^2}$.
Let
\begin{equation*}
c_j(t)=2\ell_j(k_j+\ell_j)(a_jt+b_j),\quad \gamma_j=x+[a_j-2(\ell_j+k_j^2\ell_j^{-1})]t+b_j,
\quad j=1,2,
\end{equation*}
where $a_j\in\mathbb{R}_+$, $b_j\in\mathbb{R}$ are constants.
Then one finds
\begin{equation*}
c_j(t)+\sigma(f_j,f_j)=2\ell_j(k_j+\ell_j)[\gamma_j+\rho(2k_j\ell_j)^{-1}\cos2\Theta_j],
\quad j=1,2,
\end{equation*}
\begin{equation*}
\sigma(f_1,f_2)=\rho(1+\frac{k_1-k_2}{\ell_1-\ell_2})
\cos(\Theta_1+\Theta_2)-[\rho^2-(k_1+\ell_1)(k_2+\ell_2)]
\frac{\sin(\Theta_1-\Theta_2)}{\ell_1-\ell_2}
\end{equation*}
Formulas (\ref{f40}) give a two-positon solution
\begin{subequations}
\label{f43}
\begin{equation}
q=\rho e^{2i\rho^2t}+\frac{2f_1^{(1)}f_2^{(1)}\sigma(f_1,f_2)-
2\ell_2(k_2+\ell_2)\Gamma_2(f_1^{(1)})^2-
2\ell_1(k_1+\ell_1)\Gamma_1(f_2^{(1)})^2}
{4\ell_1\ell_2(k_1+\ell_1)(k_2+\ell_2)\Gamma_1
\Gamma_2-\sigma(f_1,f_2)^2},
\end{equation}
\begin{equation}
w_1=\frac{\sqrt{2a_1\ell_1(k_1+\ell_1)}\,[2\ell_2(k_2+\ell_2)\Gamma_2
f_1^{(1)}-\sigma(f_1,f_2)f_2^{(1)}]}
{4(k_1+\ell_1)(k_2+\ell_2)\Gamma_1
\Gamma_2-\sigma(f_1,f_2)^2},
\end{equation}
\begin{equation}
w_2=\frac{\sqrt{2a_2\ell_2(k_2+\ell_2)}\,[2\ell_2(k_1+\ell_1)\Gamma_1
f_2^{(1)}-\sigma(f_1,f_2)f_1^{(1)}]}
{4(k_1+\ell_1)(k_2+\ell_2)\Gamma_1
\Gamma_2-\sigma(f_1,f_2)^2},
\end{equation}
\end{subequations}
where
\begin{equation*}
%\label{}
\Gamma_j=\gamma_j+\rho(2k_j\ell_j)^{-1}\cos2\Theta_j,\quad j=1,2.
\end{equation*}
Assume that $2\ell_1+2k_1^2\ell_1^{-1}-a_1\neq2\ell_2^2+2k_2^2\ell_2^{-1}-a_2$.
Fixing $\gamma_1$ and letting $t\rightarrow\infty$ (which implies $\gamma_2\rightarrow\infty$),
we obtain the asymptotic estimate
\begin{subequations}
\label{f44}
\begin{equation}
q=\rho e^{2i\rho^2t}-\frac{k_1\ell_1^{-1}\cos2\Theta_1-i(\sin2\Theta_1-\rho\ell_1^{-1})}
{\gamma_1+(2k_1\ell_1)^{-1}\rho\cos2\Theta_1}e^{2i\rho^2t}[1+O(t^{-1})],
\end{equation}
\begin{equation}
w_1=\sqrt{\frac{a_1}{2}}\cdot
\frac{\sqrt{1-k_1\ell_1^{-1}}\,e^{i\Theta_1}-i\sqrt{1+k_1\ell_1^{-1}}\,e^{-i\Theta_1}}
{\gamma_1+(2k_1\ell_1)^{-1}\rho\cos2\Theta_1}\,e^{i\rho^2t}[1+O(t^{-1})],
\quad
w_2=O(t^{-1}).
\end{equation}
\end{subequations}
Conversely, if we fix $\gamma_2$ and let $t\rightarrow\infty$, then we obtain
\begin{subequations}
\label{f45}
\begin{equation}
q=\rho e^{2i\rho^2t}-\frac{k_2\ell_2^{-1}\cos2\Theta_2-i(\sin2\Theta_2-\rho\ell_2^{-1})}
{\gamma_2+(2k_2\ell_2)^{-1}\rho\cos2\Theta_2}e^{2i\rho^2t}[1+O(t^{-1})],
\end{equation}
\begin{equation}
w_1=O(t^{-1}),
\quad
w_2=\sqrt{\frac{a_2}{2}}\cdot
\frac{\sqrt{1-k_2\ell_2^{-1}}\,e^{i\Theta_2}-i\sqrt{1+k_2\ell_2^{-1}}\,e^{-i\Theta_2}}
{\gamma_2+(2k_2\ell_2)^{-1}\rho\cos2\Theta_2}\,e^{i\rho^2t}[1+O(t^{-1})].
\end{equation}
\end{subequations}
Thus we have proved that the two-positon solution decays into two
positons asymptotically as $t\rightarrow\infty$, and the the collision
of the two positons are completely insensitive. Even the additional phase shifts in the
collision of two dark solitons are absent here.

\noindent\\
\textbf{(3) One-soliton-one-positon solution and soliton-positon interaction.}

We let $\rho$ satisfy $|\ell_1|<\rho<|\ell_2|$, and choose
\begin{equation*}
f_1=\left[\Psi\left(\begin{array}{c}
\sqrt{\kappa-\lambda}\,/\rho\\
0\end{array}\right)\right]_{\lambda=i\ell_1}
=\left(\begin{array}{c}
\sqrt{\kappa_1-i\ell_1}\,e^{\kappa_1(x-2\ell_1t)+i\rho^2t}\\
\sqrt{\kappa_1+i\ell_1}\,e^{\kappa_1(x-2\ell_1t)-i\rho^2t}
\end{array}\right)=
\sqrt{\rho}\,e^{\kappa_1(x-2\ell_1t)}\left(\begin{array}{c}
e^{i(\rho^2t-\theta_1)}\\e^{-i(\rho^2t-\theta_1)}
\end{array}\right),
\end{equation*}
where
\begin{equation*}
\label{}
\kappa=\sqrt{\lambda^2+\rho^2},\quad \kappa_1=\sqrt{\rho^2-\ell_1^2},\quad
\theta_1=\frac{1}{2}\arcsin\frac{\ell_1}{\rho},
\end{equation*}
and choose
\begin{equation*}
f_2=\left[\Psi\left(\begin{array}{c}
1\\-1\end{array}\right)\right]_{\lambda=i\ell_2}
=\left(\begin{array}{c}
[\rho e^{i\Theta_2}-i(k_2+\ell_2)e^{-i\Theta_2}]e^{i\rho^2t}\\
{[i(k_2+\ell_2)e^{i\Theta_2}+\rho e^{-i\Theta_2}]}e^{-i\rho^2t}
\end{array}\right),
\end{equation*}
where
\begin{equation*}
\label{}
\kappa=({\rm sign}\,{\rm Im}\lambda)\,i\sqrt{-\lambda^2-\rho^2},\quad
\Theta_2=k_2(x-2\ell_2t),\quad
k_2=({\rm sign}\,\ell_2)\sqrt{\ell_2^2-\rho^2}.
\end{equation*}
Let
\begin{equation*}
c_1(t)=\frac{\rho}{2\kappa_1}e^{2\kappa_1(a_1t+b_1)},\quad\xi_1=\kappa_1[x-(2\ell_1+a_1)t-b_1],
\end{equation*}
\begin{equation*}
c_2(t)=2\ell_2(k_2+\ell_2)(a_2t+b_2),\quad \gamma_2=x+[a_2-2(\ell_2+k_2^2\ell_2^{-1})]t+b_2,
\end{equation*}
where $a_j\in\mathbb{R}_+$, $b_j\in\mathbb{R}$, $j=1,2$, are constants.
Then one finds
\begin{equation*}
c_1(t)+\sigma(f_1,f_1)=\frac{\rho}{2\kappa_1}e^{2\kappa_1(a_1t+b_1)}(1+e^{2\xi_1}),\quad
c_2(t)+\sigma(f_2,f_2)=2\ell_2(k_2+\ell_2)[\gamma_2+(2k_2\ell_2)^{-1}\rho\cos2\Theta_2],
\end{equation*}
\begin{equation*}
\sigma(f_1,f_2)=\frac{\sqrt{\rho}\,e^{\kappa_1(x-2\ell_1t)}}{\ell_2-\ell_1}
[(k_2+\ell_2)\cos(\theta_1-\Theta_2)-\rho\sin(\theta_1+\Theta_2)].
%\frac{\rho\sin(\theta_1-\Theta_2)+(k_2+\ell_2)\cos(\theta_1+\Theta_2)
%}{\ell_2-\ell_1}e^{\kappa_1(x-2\ell_1t)}.
\end{equation*}
Formulas (\ref{f40}) give a one-soliton-one-positon solution
\begin{subequations}
\label{f46}
\begin{equation}
q=e^{2i\rho^2t}\left[\rho+\frac{2\sqrt{\rho}\,e^{2\xi_1-i\theta_1}AB-
2\ell_2(k_2+\ell_2)\rho e^{2(\xi_1-i\theta_1)}\Gamma_2-
\rho(2\kappa_1)^{-1}(1+e^{2\xi_1})A^2}
{\rho\kappa_1^{-1}\ell_2(k_2+\ell_2)(1+e^{2\xi_1})\Gamma_2-
e^{2\xi_1}B^2}\right]
\end{equation}
\begin{equation}
w_1=\frac{\sqrt{\rho a_1}\,[2\ell_2(k_2+\ell_2)\sqrt{\rho}\,e^{\xi_1-i\theta_1}
\Gamma_2-e^{\xi_1}AB]e^{i\rho^2t}}{\rho\kappa_1^{-1}\ell_2(k_2+\ell_2)
(1+e^{2\xi_1})\Gamma_2-e^{2\xi_1}B^2}
\end{equation}
\begin{equation}
w_2=\frac{\sqrt{2a_2\ell_2(k_2+\ell_2)}\,[\rho(2\kappa_1)^{-1}(1+e^{2\xi_1})A-
\sqrt{\rho}\,e^{2\xi_1-i\theta_1}B]e^{i\rho^2t}}{\rho\kappa_1^{-1}\ell_2(k_2+\ell_2)
(1+e^{2\xi_1})\Gamma_2-e^{2\xi_1}B^2}
\end{equation}
\end{subequations}
where
\begin{equation*}
\label{}
\Gamma_2=\gamma_2+\rho(2k_2\ell_2)^{-1}\cos2\Theta_2,\quad
A=%f_2^{(1)}e^{-i\rho^2t}=
\rho e^{i\Theta_2}-i(k_2+\ell_2)e^{-i\Theta_2},
\end{equation*}
\begin{equation*}
B=\frac{\sqrt{\rho}}{\ell_2-\ell_1}
[(k_2+\ell_2)\cos(\theta_1-\Theta_2)-\rho\sin(\theta_1+\Theta_2)].
\end{equation*}
Formula (\ref{f41}) implies that
\begin{equation}
\label{f46.5}
|q|^2=\rho^2-\partial_x^2\log[\rho\kappa_1^{-1}\ell_2(k_2+\ell_2)(1+e^{2\xi_1})\Gamma_2-
e^{2\xi_1}B^2].
\end{equation}
It is easy to see that
$\kappa_1^{-1}\xi_1-\gamma_2=[2(\ell_2+k_2^2\ell_2^{-1}-\ell_1)-a_1-a_2]t-b_1-b_2$.
Assume $2(\ell_2+k_2^2\ell_2^{-1}-\ell_1)-a_1-a_2>0$.
We now fix $\gamma_2$, and let $t\rightarrow-\infty$ (which implies
$\xi_1\rightarrow-\infty$), then we obtain the estimate
\begin{subequations}
\label{f47}
\begin{equation}
q=\rho e^{2i\rho^2t}+\frac{k_2\ell_2^{-1}\cos2\Theta_2-i(\sin2\Theta_2-\rho\ell^{-1})}
{\gamma_2+(2k_2\ell_2)^{-1}\rho\cos2\Theta_2}e^{2i\rho^2t}[1+O(e^{-2|\xi_1|})],
\end{equation}
\begin{equation}
w_1=O(e^{\xi_1}),\quad w_2=\sqrt{\frac{a_2}{2}}\cdot
\frac{\sqrt{1-k_2\ell_2^{-1}}\,e^{i\Theta_2}-i\sqrt{1+k_2\ell_2^{-1}}\,e^{-i\Theta_2}}
{\gamma_2+(2k_2\ell_2)^{-1}\rho\cos2\Theta_2}\,e^{i\rho^2t}[1+O(e^{-2|\xi_1|})],
\end{equation}
and
\begin{equation}
|q|^2=\rho^2-\partial_x^2\log[\gamma_2+\rho(2k_2\ell_2)^{-1}\cos2\Theta_2][1+O(e^{-2|\xi_1|})].
\end{equation}
\end{subequations}
Let $t\rightarrow+\infty$, then we obtain the estimate (for simplicity, we only give
the estimate for $|q|^2$)
\begin{equation*}
\label{f48}
|q|^2=\rho^2-\partial_x^2\log[\gamma_2+\delta_1+\rho(2k_2\ell_2)^{-1}\cos2(\Theta_2+\delta_2)]
[1+O(e^{-2|\xi_1|})],
\end{equation*}
where
\begin{equation*}
\delta_1=-\frac{\kappa_1}{\ell_2(\ell_2-\ell_1)},\quad
\delta_2=\frac{1}{2}\arcsin\frac{2\kappa_1k_2(\ell_1\ell_2-\rho^2)}{\rho^2(\ell_2-\ell_1)^2}.
\end{equation*}
If we fix $\xi_1$ and let $t\rightarrow\pm\infty$ (which implies $\gamma_2\rightarrow\pm\infty$),
then we have the asymptotic estimate
\begin{subequations}
\label{f49}
\begin{equation}
q=\rho e^{2i\rho^2t}-\frac{1+e^{-4i\theta_1}}{1+e^{2\xi_1}}e^{2\xi_1+2i\rho^2t}[1+O(t^{-1})],
\end{equation}
\begin{equation}
w_1=\frac{2\sqrt{a_1}\,\kappa_1e^{\xi_1-i\theta_1}}
{1+e^{2\xi_1}}e^{i\rho^2t}[1+O(t^{-1})],
\quad
w_2=O(t^{-1}).
\end{equation}
\end{subequations}
Thus we have proved that the one-soliton-one-positon solution decays asymptotically into
a dark soliton and a positon for large $t$. The dark soliton recovers completely
after the collision with a positon, in other words, a positon is totally transparent to
a dark soliton. However, the positon gains phase shifts when colliding with the dark
soliton.

In Figure 2, we plot the one-soliton-one-positon solution.
\begin{figure}[htp]
\centering
\includegraphics[width=2in,angle=270]{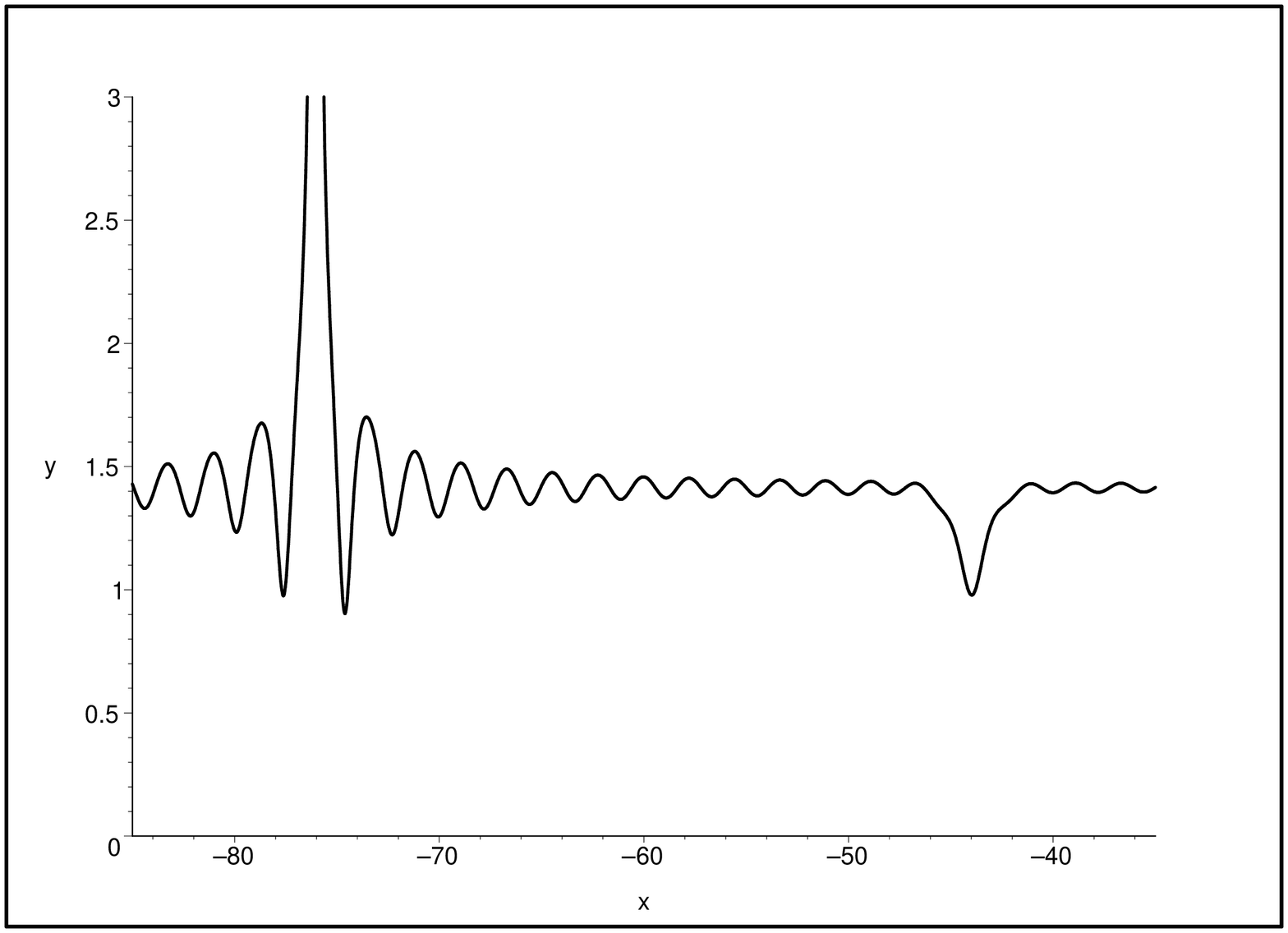}
\includegraphics[width=2in,angle=270]{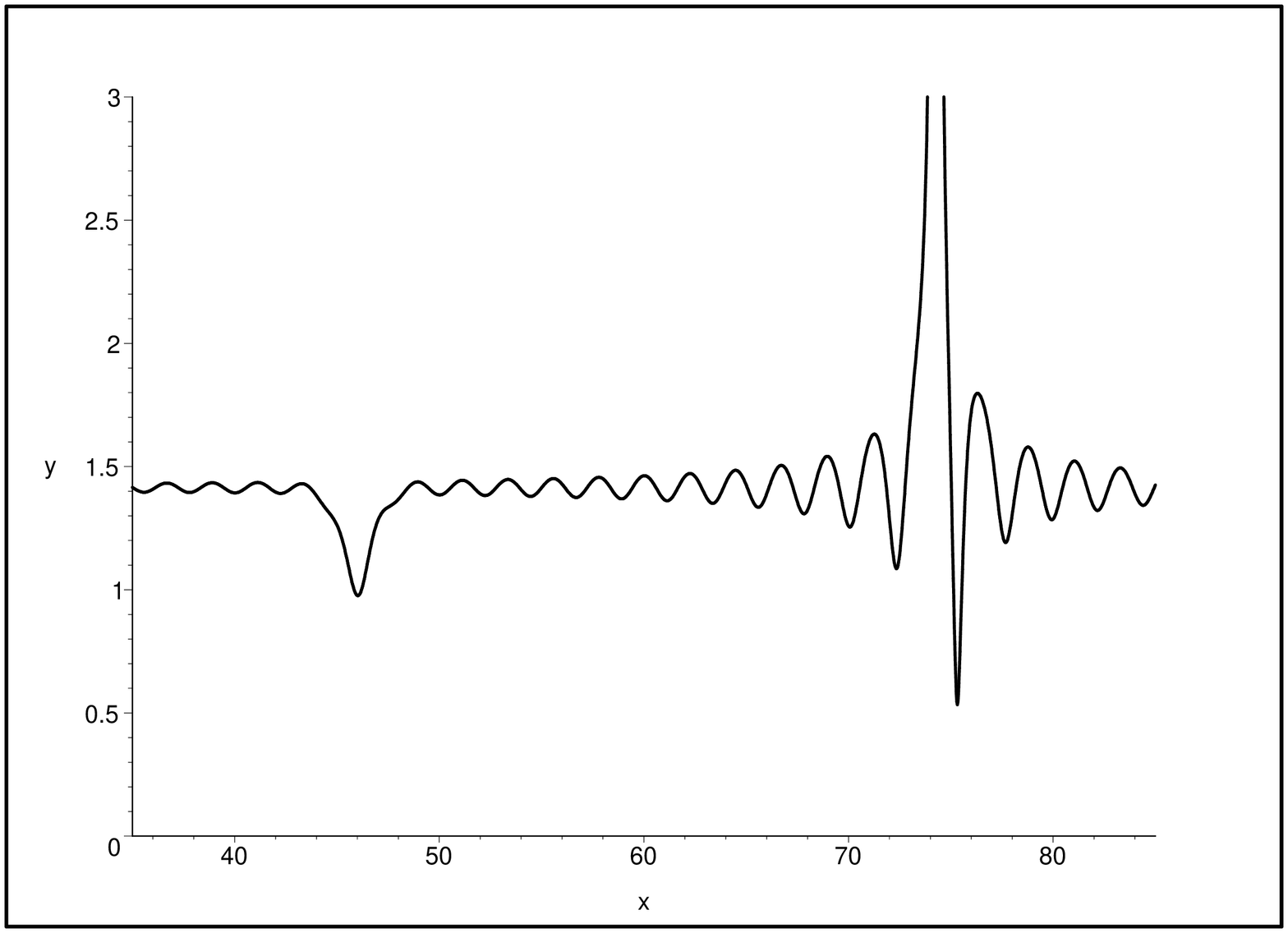}
\caption{{\footnotesize The one-soliton-one-positon solution of the NLS$^+$ESCS
(\ref{f39}) with $\ell_1=1$ and $\ell_2=2$. The data is $\rho=\sqrt{2}$ and
$a_1=a_2=b_1=b_2=1$. The two graphs show the modulus of $q$ at $t=-15$ (the left)
and $t=15$ (the right) respectively.}}
\end{figure}
\subsubsection{Solutions of the NLS$^+$ESCS with $m=0$ and $n=N$.}
The NLS$^+$ESCS with $m=0$ and $n=N$ reads
\begin{subequations}
\label{f50}
\begin{equation}
w_{j,x}=i\ell_jw_j+q w^*_j,\quad
j=1,\ldots,N,
\end{equation}
\begin{equation}
q_t=i(2|q|^2q-q_{xx})+\sum_{j=1}^N w_j^2,
\end{equation}
\end{subequations}
where $\ell_j\neq0$, $j=1,\ldots,N$, are $N$ distinct real constants.
For $j=1,\ldots,N$, let $f_j$ be a solution of the system (\ref{f4}) with $\lambda=i\ell_j$
and satisfy $f_j^{(1)}=f_j^{(2)*}$, and let $c_j(t)$ be an arbitrary real function satisfying
$\dot{c}_j(t)\geq0$. Then by Proposition \ref{pr.e3}, a solution of the equations (\ref{f50})
is given by
%\begin{subequations}
%\label{f51}
\begin{equation}
\label{f51}
q=\rho e^{2i\rho^2t}+\frac{\Delta_2}{\Delta_0},\quad
%end{equation}
%\begin{equation}
%\label{f51b}
w_j=\frac{\sqrt{\dot{c}_j(t)}\,\Delta_{1j}}{\Delta_0},\quad j=1,\ldots,N,
\end{equation}
%\end{subequations}
where
\begin{equation*}
\label{}
\Delta_0=W_0(\{c_1,f_1\},\ldots,\{c_N,f_N\}),\quad
\Delta_2=W_2^{(1)}(\{c_1,f_1\},\ldots,\{c_N,f_N\};0),
\end{equation*}
\begin{equation*}
\label{}
\Delta_{1j}=W_1^{(1)}(\{c_1,f_1\},\ldots,\{c_{j-1},f_{j-1}\},\{c_{j+1},f_{j+1}\},
\ldots,\{c_N,f_N\};f_j).
\end{equation*}
Moreover, we have
\begin{equation}
\label{}
|q|^2=\rho^2-\partial_x^2\log\Delta_0.
\end{equation}
For simplicity, we assume $|\ell_1|>\ldots>|\ell_N|$. Then according to the different
choice of $\rho$, we can obtain different classes of solutions.

\noindent\\
\textbf{(1) Multi-soliton solutions.}

We take $\rho>|\ell_j|$, $j=1,\ldots,N$, and choose
\begin{equation*}
\label{}
c_j(t)=\frac{\rho}{2\kappa_j}e^{2\kappa_j(a_jt+b_j)},\quad
f_j=\left[\Psi\left(\begin{array}{c}
\sqrt{\kappa-\lambda}\,/\rho\\0
\end{array}\right)\right]_{\lambda=i\ell_j}
=\left(\begin{array}{c}
\sqrt{\kappa_j-i\ell_j}\,e^{\kappa_j(x-2\ell_jt)+i\rho^2t}\\
\sqrt{\kappa_j+i\ell_j}\,e^{\kappa_j(x-2\ell_jt)-i\rho^2t}
\end{array}\right),
\end{equation*}
where $\kappa=\sqrt{\lambda^2+\rho^2}$, $\kappa_j=\sqrt{\rho^2-\ell_j^2}$,
$a_j\in\mathbb{R}_+$ and $b_j\in\mathbb{R}$,
then formulas (\ref{f51}) give the dark $N$-soliton solution.

\noindent\\
\textbf{(2) Multi-positon solutions.}

We take $\rho<|\ell_j|$, $j=1,\ldots,N$, and choose
\begin{equation*}
c_j(t)=2\ell_j(k_j+\ell_j)(a_jt+b_j),\quad
f_j=\left[\Psi\left(\begin{array}{c}
1\\-1\\
\end{array}\right)\right]_{\lambda=i\ell_j}=
\left(\begin{array}{c}
[\rho e^{i\Theta_j}-i(k_j+\ell_j)e^{-i\Theta_j}]e^{i\rho^2t}\\
{[i(k_j+\ell_j)e^{i\Theta_j}+\rho e^{-i\Theta_j}]}e^{-i\rho^2t}
\end{array}\right),
\end{equation*}
where $\kappa=({\rm sign}\,{\rm Im}\lambda)\,i\sqrt{-\lambda^2-\rho^2}$,
$k_j=({\rm sign}\,\ell_j)\sqrt{\ell_j^2-\rho^2}$,
$a_j\in\mathbb{R}_+$ and $b_j\in\mathbb{R}$, then formulas (\ref{f51})
give the $N$-positon solution.

\noindent\\
\textbf{(3) Multi-soliton-multi-positon solutions.}

We let $\rho$ satisfy $|\ell_{N_1}|>\rho>|\ell_{N_1+1}$, where $1\leq N_1\leq N$,
and choose
\begin{equation*}
\label{}
c_j(t)=\frac{\rho}{2\kappa_j}e^{2\kappa_j(a_jt+b_j)},\quad
f_j=\left[\Psi\left(\begin{array}{c}
\sqrt{\kappa-\lambda}\,/\rho\\0
\end{array}\right)\right]_{\lambda=i\ell_j},
\quad j=1,\ldots,N_1,
\end{equation*}
where $\kappa=\sqrt{\lambda^2+\rho^2}$ and $\kappa_j=\sqrt{\rho^2-\ell_j^2}$,
and
\begin{equation*}
c_j(t)=2\ell_j(k_j+\ell_j)(a_jt+b_j),\quad
f_j=\left[\Psi\left(\begin{array}{c}
1\\-1\\
\end{array}\right)\right]_{\lambda=i\ell_j}
, \quad j=N_1+1,\ldots,N,
\end{equation*}
where $\kappa=({\rm sign}\,{\rm Im}\lambda)\,i\sqrt{-\lambda^2-\rho^2}$ and
$k_j=({\rm sign}\,\ell_j)\sqrt{\ell_j^2-\rho^2}$.
Here $a_j\in\mathbb{R}_+$ and $b_j\in\mathbb{R}$ for $j=1,\ldots,N$.
Then formulas (\ref{f51}) give the $N_1$-soliton-$N_2$-positon solution ($N_2=N-N_1$).

\subsection{Solutions of the NLS$^-$ESCS}
\hskip\parindent
We start from the NLS$^-$ equation without sources
\begin{equation}
\label{f52}
q_t=-i(2|q|^2q+q_{xx}),
\end{equation}
and its solution
\begin{equation}
\label{f53}
q=\rho e^{-2i\rho^2t}.
\end{equation}
We need to solve the linear system
\begin{equation}
\label{f54}
\psi_x=U(\lambda,\rho e^{-2i\rho^2t},-\rho e^{2i\rho^2t})\psi,
\quad \psi_t=V(\lambda,\rho e^{-2i\rho^2t},-\rho e^{2i\rho^2t})\psi.
\end{equation}
The fundamental solution matrix for the linear system (\ref{f54}) is
\begin{equation}
\label{f55}
\Phi=\left(\begin{array}{cc}
(\kappa+\lambda)e^{\kappa(x+2i\lambda t)-i\rho^2t}&-\rho e^{-\kappa(x+2i\lambda t)-i\rho^2t}\\
-\rho e^{\kappa(x+2i\lambda t)+i\rho^2t}&(\kappa+\lambda)e^{-\kappa(x+2i\lambda t)+i\rho^2t}
\end{array}\right),
\end{equation}
where $\kappa=\kappa(\lambda)$ satisfies $\kappa^2=\lambda^2-\rho^2$.

\subsubsection{Solutions of the NLS$^-$ESCS with $n=1$.}
\hskip\parindent
The NLS$^-$ESCS with $n=1$ reads
\begin{subequations}
\label{f56}
\begin{equation}
\label{f56a}
\varphi_{1,x}=U(\lambda_1,q,-q^*)\varphi_1,
\end{equation}
\begin{equation}
\label{f56b}
q_t=-i(2|q|^2q+q_{xx})+(\varphi_1^{(1)})^2-(\varphi_1^{(2)*})^2,
\end{equation}
\end{subequations}
where $\lambda_1=\lambda_{1R}+i\lambda_{1I}$ is a complex constant with $\lambda_{1R}>0$,
$\lambda_{1I}\neq0$.
Let $f$ be a solution of the system (\ref{f54}) with $\lambda=\lambda_1$,
$c(t)$ be an arbitrary complex function, then by Proposition \ref{pr.e4},
a solution of the equations (\ref{f56}) is given by
\begin{equation}
\label{f57}
q=\rho e^{-2i\rho^2t}+\frac{\Delta_2}{\Delta_0},
\quad
\varphi_1=\frac{\sqrt{\dot{c}(t)}}{\Delta_0}\left(\begin{array}{c}
\Delta_1^{(1)}\\\Delta_1^{(2)}\end{array}\right),
\end{equation}
where
\begin{equation*}
\label{}
\Delta_0=\left|\begin{array}{cc}
c(t)+\sigma(f,f)&-\frac{|f^{(1)}|^2+|f^{(2)}|^2}{4\lambda_{1R}}\\
-\frac{|f^{(1)}|^2+|f^{(2)}|^2}{4\lambda_{1R}}&-c(t)^*-\sigma(f,f)^*
\end{array}\right|=-|c(t)+\sigma(f,f)|^2-
\left(\frac{|f^{(1)}|^2+|f^{(2)}|^2}{4\lambda_{1R}}\right)^2,
\end{equation*}
\begin{equation*}
\label{}
\Delta_1^{(1)}=\left|\begin{array}{cc}
-c(t)^*-\sigma(f,f)^*&-\frac{|f^{(1)}|^2+|f^{(2)}|^2}{4\lambda_{1R}}\\
-f^{(2)*}&f^{(1)}
\end{array}\right|,
\quad
\Delta_1^{(2)}=\left|\begin{array}{cc}
-c(t)^*-\sigma(f,f)^*&-\frac{|f^{(1)}|^2+|f^{(2)}|^2}{4\lambda_{1R}}\\
f^{(1)*}&f^{(2)}
\end{array}\right|,
\end{equation*}
\begin{equation*}
\Delta_2=\left|\begin{array}{ccc}
c(t)+\sigma(f,f)&-\frac{|f^{(1)}|^2+|f^{(2)}|^2}{4\lambda_{1R}}&f^{(1)}\\
-\frac{|f^{(1)}|^2+|f^{(2)}|^2}{4\lambda_{1R}}&-c(t)^*-\sigma(f,f)^*&-f^{(2)*}\\
f^{(1)}&-f^{(2)*}&0
\end{array}\right|.
\end{equation*}
Moreover, we have
\begin{equation}
\label{f58}
|q|^2=\rho^2+\partial_x^2\log\Delta_0
\end{equation}
\noindent
\textbf{Topological deformation of the bright one-soliton.}

We choose $f$ as
\begin{equation*}
\label{}
f=\left[\Phi\left(\begin{array}{c}
1\\0\end{array}\right)\right]_{\lambda=\lambda_1}
=\left(\begin{array}{c}
(\kappa_1+\lambda_1)e^{-i\rho^2t}\\
-\rho e^{i\rho^2t}
\end{array}\right)
e^{\kappa_1(x+2i\lambda_1 t)},
\end{equation*}
where $\kappa_1=\kappa(\lambda_1)$.
Here, we choose $\kappa=\kappa(\lambda)=({\rm sign}\,\lambda_I)\sqrt{\lambda^2-\rho^2}$ for $\Phi$
defined by (\ref{f55}), then $\kappa$ is %analytic in the domain
%$\mathbb{C}\backslash(\{z||z_R|=0\}\cup\{z||z_R|\leq\rho, |z_I|=0\})$, and hence is
analytic at $\lambda=\lambda_1$. Furthermore, under this choice of $\kappa$,
we have $\lim_{\rho\rightarrow0}\kappa=\lambda$. Calculation yields
\begin{equation*}
\sigma(f,f)=\frac{\rho(\kappa_1+\lambda_1)}{2\kappa_1}e^{2\kappa_1(x+2i\lambda_1 t)},
\quad |f^{(1)}|^2=|\kappa_1+\lambda_1|^2e^{2(\kappa_{1R}x-2\lambda_{1I}t)},
\quad |f^{(2)}|^2=\rho^2e^{2(\kappa_{1R}x-2\lambda_{1I}t)}.
\end{equation*}
We choose $c(t)=(2\kappa_1)^{-1}(\kappa_1+\lambda_1)e^{2(at+b)}$,
where $a$ and $b$ are two arbitrary complex numbers, then formulas (\ref{f57}) give the
topological deformation of bright one-soliton solution
\begin{subequations}
\label{f59}
\begin{equation}
\label{f59a}
q=\left[\rho+\frac{\frac{\rho^2(\kappa_1+\lambda_1)}{2\kappa_1}e^{-2i\eta}
-\frac{|\kappa_1+\lambda_1|^2(\kappa_1+\lambda_1)}{2\kappa_1^*}e^{2i\eta}
+\frac{\rho(\kappa_1+\lambda_1)}{2}(\frac{\rho^2}{\kappa_1}+\frac{|\kappa_1+\lambda_1|^2
+\rho^2}{\lambda_{1R}}-\frac{|\kappa_1+\lambda_1|^2}{\kappa_1^*})e^{2\xi}}
{\frac{|\kappa_1+\lambda_1|^2}{4|\kappa_1|^2}(e^{-2\xi}
+2\rho\cos2\eta+\rho^2e^{2\xi})+\left(\frac{|\kappa_1+\lambda_1|^2+\rho^2}{4\lambda_{1R}}
\right)^2e^{2\xi}}\right]e^{-2i\rho^2t},
\end{equation}
\begin{equation}
\label{f59b}
\varphi_1^{(1)}=\sqrt{\frac{a(\kappa_1+\lambda_1)}{\kappa_1}}\cdot
\frac{\frac{|\kappa_1+\lambda_1|^2}{2\kappa_1^*}(e^{-\xi+i\eta}+\rho e^{\xi-i\eta})
-\frac{\rho(|\kappa_1+\lambda_1|^2+\rho^2)}{4\lambda_{1R}}e^{\xi-i\eta}}
{\frac{|\kappa_1+\lambda_1|^2}{4|\kappa_1|^2}(e^{-2\xi}
+2\rho\cos2\eta+\rho^2e^{2\xi})+\left(\frac{|\kappa_1+\lambda_1|^2+\rho^2}{4\lambda_{1R}}
\right)^2e^{2\xi}}e^{-i\rho^2t},
\end{equation}
\begin{equation}
\label{f59c}
\varphi_1^{(2)}=\sqrt{\frac{a(\kappa_1+\lambda_1)}{\kappa_1}}\cdot
\frac{\frac{-\rho(\kappa_1^*+\lambda_1^*)}{2\kappa_1^*}(e^{-\xi+i\eta}+\rho e^{\xi-i\eta})
-\frac{(\kappa_1^*+\lambda_1^*)(|\kappa_1+\lambda_1|^2+\rho^2)}{4\lambda_{1R}}e^{\xi-i\eta}}
{\frac{|\kappa_1+\lambda_1|^2}{4|\kappa_1|^2}(e^{-2\xi}
+2\rho\cos2\eta+\rho^2e^{2\xi})+\left(\frac{|\kappa_1+\lambda_1|^2+\rho^2}{4\lambda_{1R}}
\right)^2e^{2\xi}}e^{i\rho^2t},
\end{equation}
\end{subequations}
where
\begin{equation*}
\xi=\kappa_{1R}x-(2\lambda_{1I}+a_R)t-b_R,\quad
\eta=\kappa_{1I}x+(2\lambda_{1R}-a_I)t-b_I.
\end{equation*}
Formula (\ref{f58}) implies that
\begin{equation*}
|q|^2=\rho^2+\partial_x^2\log\left[4\lambda_{1R}^2|\kappa_1+\lambda_1|^2(e^{-2\xi}
+2\rho\cos2\eta+\rho^2e^{2\xi})+|\kappa_1|^2(|\kappa_1+\lambda_1|^2+\rho^2)^2e^{2\xi}\right].
\end{equation*}
When $\rho=0$, we have $\kappa_1=\lambda_1$ and the solution given by (\ref{f59})
corresponds to the bright one-soliton solution
\begin{equation*}
q=-\frac{2\lambda_{1R}\,e^{2i\eta_0}}{\cosh2\xi_0},\quad
\varphi_1=\frac{\sqrt{2a\lambda_{1R}}}{\cosh2\xi_0}\left(\begin{array}{c}
e^{-\xi_0+i\eta_0}\\-e^{\xi_0-i\eta_0}\end{array}\right),
\end{equation*}
where
\begin{equation*}
\xi_0=\lambda_{1R}x-(2\lambda_{1I}+a_R)t-b_R+\log(|\lambda_1|/\sqrt{\lambda_{1R}}),
\quad \eta_0=\lambda_{1I}+(2\lambda_{1R}-a_I)t-b_I+\arg\lambda_1.
\end{equation*}
The topological deformation of
bright one-soliton solution for the NLS$^-$ equation was already known. Here, we have
given its correspondence for the NLS$^-$ESCS.

In Figure 3, we plot the topological deformation of bright one-soliton solution.
\begin{figure}[htp]
\centering
\includegraphics[width=2in,angle=270]{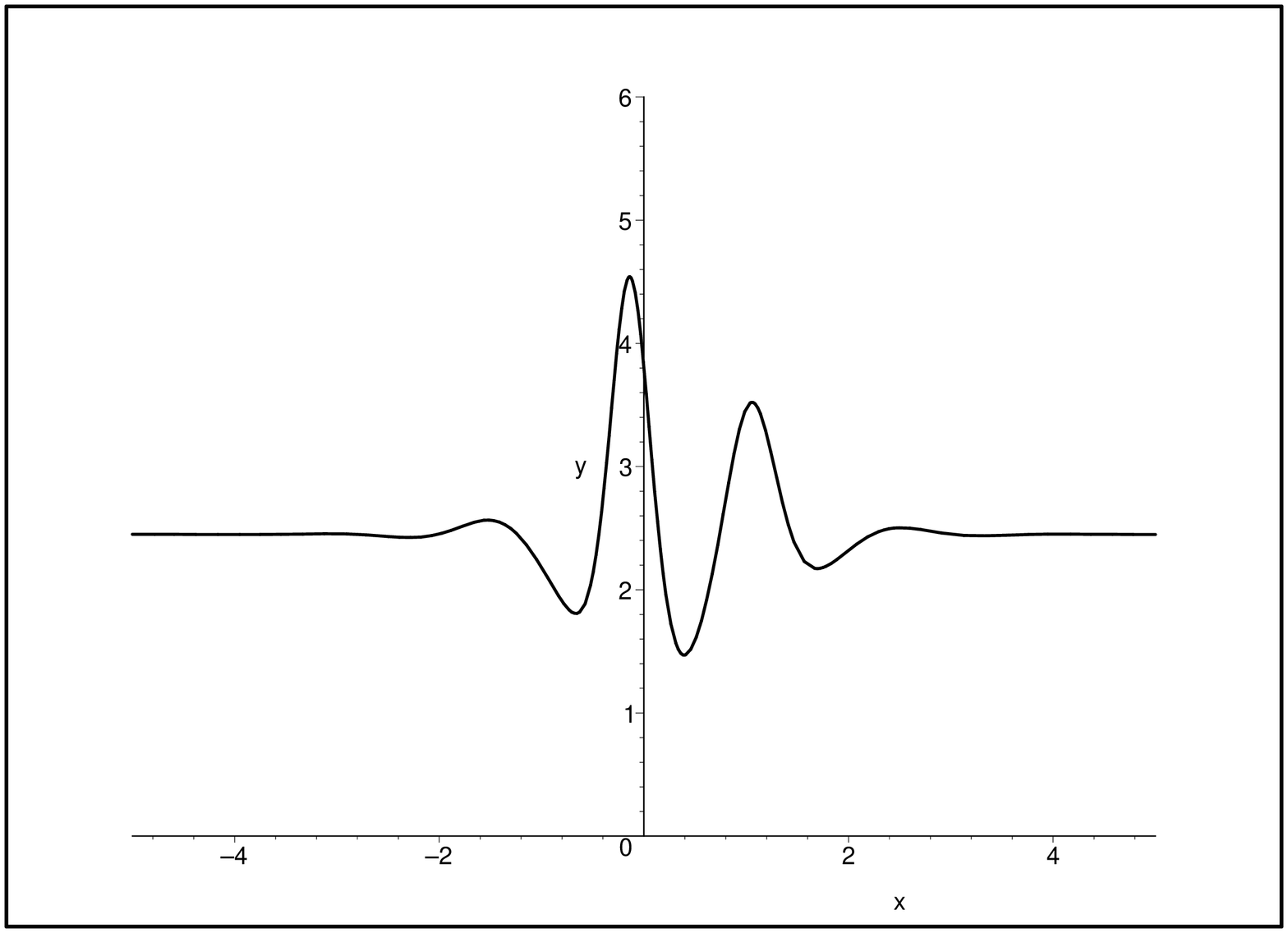}
\includegraphics[width=2in,angle=270]{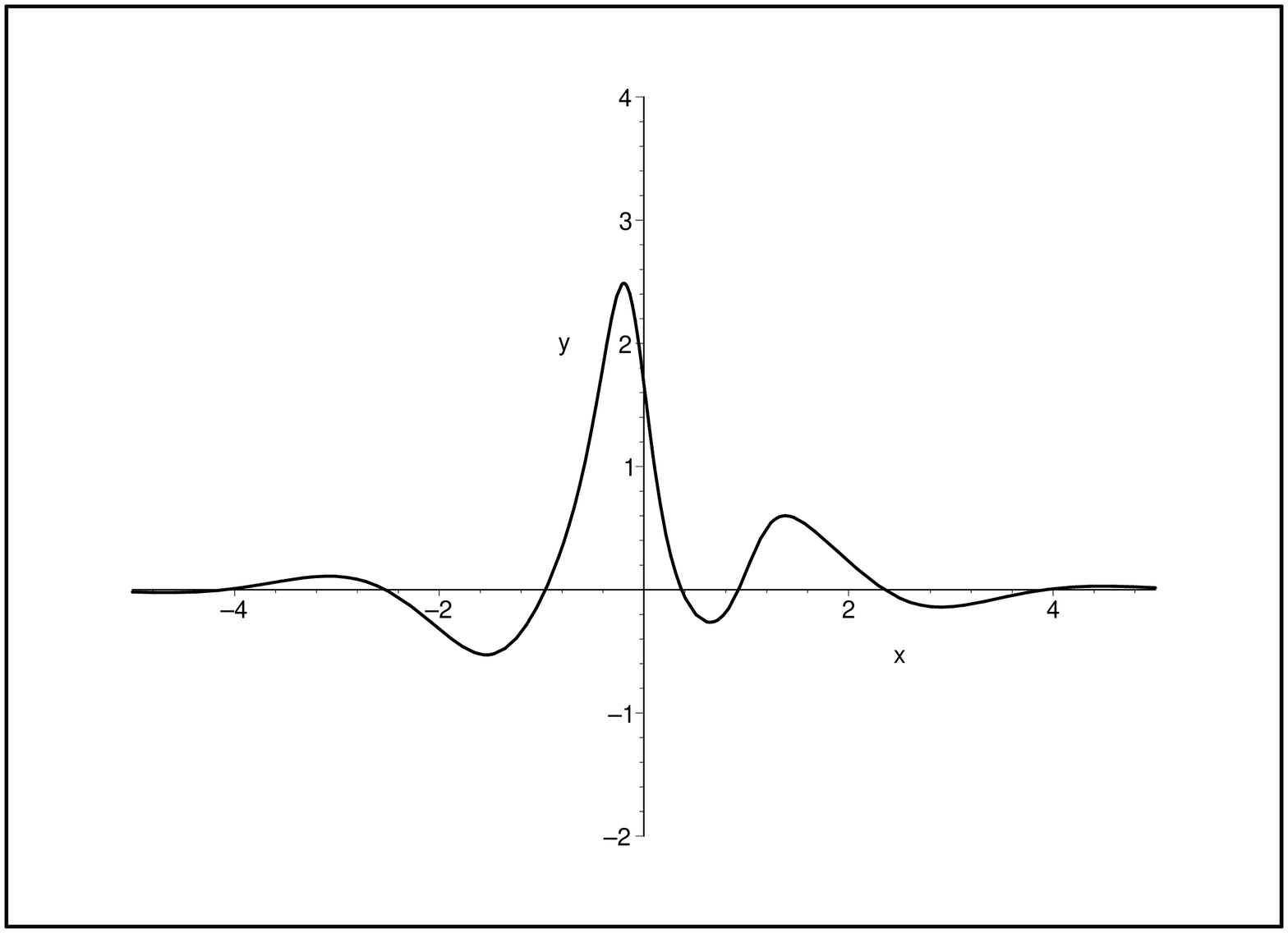}
\caption{{\footnotesize The topological deformation of bright one-soliton solution of
the NLS$^-$ESCS (\ref{f56}) with $\lambda_1=2+i$. The data is $\rho=\sqrt{6}$ and
$a_1=a_2=b_1=b_2=1$. The two graphs show the modulus of $q$ (the left) and the real part of
$\phi_1^{(1)}$ (the right) at $t=0$.}}
\end{figure}
\subsubsection{Solutions of the NLS$^-$ESCS with $n=N$.}
\hskip\parindent
The NLS$^-$ESCS with $n=N$ reads
\begin{subequations}
\label{f60}
\begin{equation}
\label{f60a}
\varphi_{j,x}=U(\lambda_j,q,-q^*)\varphi_j,\quad j=1,\ldots,N,
\end{equation}
\begin{equation}
\label{f60b}
q_t=-i(2|q|^2q+q_{xx})+(\varphi_1^{(1)})^2-(\varphi_1^{(2)*})^2,
\end{equation}
\end{subequations}
where $\lambda_j=\lambda_{jR}+i\lambda_{jI}$ are distinct complex constants with
$\lambda_{jR}>0$, $\lambda_{jI}\neq0$. For $j=1,\ldots,N$, let
\begin{equation*}
F_j=\{c_j,f_j\},\ F_j'=\{-c_j^*,S_-f_j\},\
c_j(t)=\frac{\kappa_j+\lambda_j}{2\kappa_j}e^{a_jt+b_j},\
f_j=\left[\Phi\left(\begin{array}{c}
1\\0\end{array}\right)\right]_{\lambda=\lambda_j},\
\kappa_j=({\rm sign}\,\lambda_{jI})\sqrt{\lambda_j^2-\rho^2},
\end{equation*}
then the topological deformation of bright $N$-soliton solution of Eqs. (\ref{f60}) is given by
\begin{equation*}
q=\rho e^{-2i\rho^2t}+\frac{\Delta_2}{\Delta_0},\quad
\varphi_j=\frac{\sqrt{\dot{c}_j(t)}}{\Delta_0}\left(\begin{array}{c}
\Delta_{1j}^{(1)}\\\Delta_{1j}^{(2)}\end{array}\right),\quad j=1,\ldots,N,
\end{equation*}
where
\begin{equation*}
\Delta_0=W_0(F_1,F_1',\ldots,F_N,F_N'),\quad
\Delta_2=W_2^{(0)}(F_1,F_1',\ldots,F_N,F_N';0),
\end{equation*}
\begin{equation*}
\Delta_{1j}^{(l)}=W_1^{(l)}(F_1,F_1',\ldots,F_{j-1},F_{j-1}',F'_j,F_{j+1},F_{j+1}',
\ldots,F_N,F_N';f_j),\quad l=1,2,\ j=1,\ldots,N.
\end{equation*}

\section*{ Acknowledgment}\hskip\parindent
This work was supported by the Chinese Basic Research Project
"Nonlinear Science".

\end{document}